\documentclass[twocolumn, twocolappendix]{aastex631}

\usepackage{graphicx, graphics}
\usepackage{CJK}
\usepackage{bbm}
\usepackage{amsmath}
\usepackage{bm}
\usepackage{comment}
\usepackage{colortbl}
\usepackage{xcolor}
\usepackage{booktabs}
\usepackage{makecell}
\usepackage[normalem]{ulem}
\definecolor{myblue}{HTML}{1f77b4}

\begin{document}
\begin{CJK*}{UTF8}{gkai}

\title{Complex Interplay between Outer Disk Dynamics and Inner Water Reservoir}

\author[0000-0002-7607-719X]{Feng Long (龙凤)}
\altaffiliation{NASA Hubble Fellowship Program Sagan Fellow}
\affiliation{Kavli Institute for Astronomy and Astrophysics, Peking University, Beijing 100871, China}
\affiliation{Department of Astronomy, Peking University, Beijing 100871, China}
\affiliation{Lunar and Planetary Laboratory, University of Arizona, Tucson, AZ 85721, USA}
\email{long.feng@pku.edu.cn}

\author[0000-0001-7962-1683]{Ilaria Pascucci}
\affiliation{Lunar and Planetary Laboratory, University of Arizona, Tucson, AZ 85721, USA}

\author[0000-0001-8764-1780]{Paola Pinilla}
\affiliation{Mullard Space Science Laboratory, University College London, Holmbury St Mary, Dorking, Surrey RH5 6NT, UK.}

\author[0000-0003-4335-0900]{Andrea Banzatti}
\affil{Department of Physics, Texas State University, 749 N Comanche Street, San Marcos, TX 78666, USA}

\author[0000-0001-8790-9011]{Adrien Houge}
\affiliation{Center for Star and Planet Formation, GLOBE Institute, University of Copenhagen, {\O}ster Voldgade 5-7, DK-1350 Copenhagen, Denmark}

\author[0000-0003-2253-2270]{Sean M. Andrews}
\affiliation{Center for Astrophysics \textbar\, Harvard \& Smithsonian, 60 Garden St., Cambridge, MA 02138, USA}

\author[0000-0002-6773-459X]{Doug Johnstone}
\affiliation{NRC Herzberg Astronomy and Astrophysics, 5071 West Saanich Rd, Victoria, BC, V9E 2E7, Canada}
\affiliation{Department of Physics and Astronomy, University of Victoria, Victoria, BC, V8P 5C2, Canada}

\author[0000-0002-7154-6065]{Gregory J. Herczeg (沈雷歌)}
\affiliation{Kavli Institute for Astronomy and Astrophysics, Peking University, Beijing 100871, China}
\affiliation{Department of Astronomy, Peking University, Beijing 100871, China}

\author[0000-0002-5758-150X]{Joan Najita}
\affiliation{NSF NOIRLab, 950 N. Cherry Avenue, Tucson, AZ 85719, USA}

\author[0000-0001-7552-1562]{Klaus M. Pontoppidan}
\affiliation{Jet Propulsion Laboratory, California Institute of Technology, 4800 Oak Grove Drive, Pasadena, CA 91109, USA}

\author[0009-0008-8176-1974]{Joe Williams}
\affiliation{Department of Physics and Astronomy, University of Exeter, Exeter, EX4 4QL, UK}

\author[0000-0002-0661-7517]{Ke Zhang}
\affiliation{Department of Astronomy, University of Wisconsin-Madison,
475 N Charter St, Madison, WI 53706, USA}

\author[0000-0003-0787-1610]{Geoffrey A. Blake}
\affiliation{Division of Geological and Planetary Sciences, California Institute of Technology, MC 150-21, Pasadena, CA 91125, USA}

\author[0000-0002-5296-6232]{Mar\'{i}a Jos\'{e} Colmenares}
\affiliation{Department of Astronomy, University of Michigan, Ann Arbor, MI 48109, USA}

\author[0000-0003-4853-5736]{Giovanni Rosotti}
\affiliation{Dipartimento di Fisica, Universit\`a degli Studi di Milano, Via Celoria 16, 20133 Milano, Italy}

\author[0000-0001-5378-7749]{Enrico Ragusa}
\affiliation{Dipartimento di Fisica, Universit\`a degli Studi di Milano, Via Celoria 16, 20133 Milano, Italy}

\author[0000-0003-3682-6632]{Colette Salyk}
\affiliation{Department of Physics and Astronomy, Vassar College, 124 Raymond Avenue, Poughkeepsie, NY 12604, USA}

\author[0000-0001-8184-5547]{Chengyan Xie}
\affiliation{Lunar and Planetary Laboratory, University of Arizona, Tucson, AZ 85721, USA}

\author[0000-0003-3190-8890]{Tayt Armitage}
\affiliation{Department of Astronomy, University of Wisconsin-Madison,
475 N Charter St, Madison, WI 53706, USA}

\author[0000-0001-6906-9549]{Xue-Ning Bai}
\affiliation{Institute for Advanced Study and Department of Astronomy, Tsinghua University, 100084, Beijing, China}

\author[0000-0001-6307-4195]{Daniel Harsono}
\affiliation{Institute of Astronomy, Department of Physics, National Tsing Hua University, Hsinchu, Taiwan}

\author[0000-0001-8240-978X]{Till Kaeufer}
\affiliation{Department of Physics and Astronomy, University of Exeter, Exeter, EX4 4QL, UK}

\author[0000-0002-3291-6887]{Sebastiaan Krijt}
\affiliation{Department of Physics and Astronomy, University of Exeter, Exeter, EX4 4QL, UK}

\author[0000-0002-2357-7692]{Giuseppe Lodato}
\affiliation{Dipartimento di Fisica, Universit\`a degli Studi di Milano, Via Celoria 16, 20133 Milano, Italy}

\author[0000-0002-7616-666X]{Yao Liu}
\affiliation{School of Physical Science and Technology, Southwest Jiaotong University, Chengdu 610031, China}

\author[0000-0003-3562-262X]{Carlo F. Manara}
\affiliation{European Southern Observatory, Karl-Schwarzschild-Str. 2, D-85748 Garching bei M\"{u}nchen, Germany}

\author[0000-0002-1637-7393]{Fran\c{c}ois M\'enard}
\affiliation{Univ.\ Grenoble Alpes, CNRS, IPAG, 38000 Grenoble, France}

\author[0000-0002-0554-1151]{Mayank Narang}
\affiliation{Jet Propulsion Laboratory, California Institute of Technology, 4800 Oak Grove Drive, Pasadena, CA 91109, USA}

\author[0000-0002-4147-3846]{Miguel Vioque}
\affiliation{European Southern Observatory, Karl-Schwarzschild-Str. 2, D-85748 Garching bei M\"{u}nchen, Germany}

\author[0000-0003-1526-7587]{David J. Wilner}
\affiliation{Center for Astrophysics \textbar\, Harvard \& Smithsonian, 60 Garden St., Cambridge, MA 02138, USA}

\author{the JDISCS collaboration}

\begin{abstract}

We present a homogeneous characterization of outer disk dust and gas properties for 51 protoplanetary disks from the JWST/MIRI Cycle 1 programs, combining new ALMA observations with archival data. Visibility fitting with \texttt{frank} recovers more and higher-contrast substructures in the mm continuum than image-plane analysis, placing the innermost features of several disks at 4--5\,au and leaving only three disks featureless at the current resolution. CO gas disk sizes span $\sim$60--1000\,au and correlate strongly with mm continuum sizes, with $R_{\rm CO}/R_{\rm mm}$ ranging from 1.4 to 9.7 (median 2.95).
We then connect these outer disk properties to JWST/MIRI inner disk water diagnostics. We find no global correlation between the cold-to-hot water flux ratio ($F_{\rm cold}/F_{\rm hot}$), a proposed proxy of cold water enrichment by pebble transport, and any single outer disk parameter, including $R_{\rm mm}$, $R_{\rm CO}/R_{\rm mm}$, and innermost substructure location. Three compact ($R_{\rm mm}<30$\,au), structureless disks with $R_{\rm CO}/R_{\rm mm}>4$ show both elevated $F_{\rm cold}/F_{\rm hot}$ and cold water emission near the snowline temperature, consistent with unimpeded inward drift of icy pebbles. For structured disks, the water properties instead vary widely at fixed gap location, reflecting the sensitivity of the inner disk water content to multiple factors, including gap leakiness, grain size, and thermal history. Linking outer disk dust evolution to inner disk water content thus requires carefully selected samples with detailed constraints on these additional parameters.

\end{abstract}


\section{Introduction} \label{sec:intro}
The composition of planets is set, in part, by the volatile budget available in the protoplanetary disk during the epoch of planet formation. Within the inner few AU of the protoplanetary disk, gas-phase molecules including H$_2$O, CO$_2$, HCN, and C$_2$H$_2$ trace the local chemical environment, where rocky planets assemble. Water is of particular interest, both as one of the most abundant volatile species and as a tracer of the delivery of oxygen-rich icy material to the inner disk \citep{vanDishoeck2021}. Understanding how this inner disk water reservoir is built up and replenished over time is therefore central to the broader question of how planets acquire their volatile inventories.

Infrared spectroscopy provides the main window into this reservoir. Spitzer/IRS observations first established that water vapor and simple organics are common in the inner region of T Tauri disks \citep{Carr2008, Salyk2008, Pontoppidan2010}. Water has a dense forest of rotational transitions across mid-infrared wavelengths, with upper-level energies ($E_{\rm up}$) from a few hundred to several thousand Kelvin, that together probe different regions of the disk, from the hot innermost AU to the cooler gas near the water snowline. Retrieving this radial information is essential as it ties the observed emission to the physical processes acting at each location \citep{Banzatti2023a}. Spitzer/IRS could not access it, however, because its resolving power ($R\sim600$--700) left transitions blended and slab-model fits carry large uncertainties in derived gas properties \citep{Salyk2011}.
The Medium Resolution Spectrometer \citep[MRS; $R\sim1800$--4000,][]{Wells2015} on JWST/MIRI \citep{Rieke2015} now separates many individual lines, and curated lists of unblended transitions allow water excitation conditions to be characterized across the disk population \citep{Banzatti2025, Gasman2025}. These advances make it possible to robustly test what physically regulates the inner disk water reservoir.

One leading hypothesis is that radial pebble drift from the outer disk plays a central role. Beyond the water snowline ($\sim$1--3\,AU in the midplane for young Sun-like stars), water vapor condenses onto dust grains, forming icy pebbles that populate the cold outer disk. Aerodynamic drag drives these pebbles inward \citep{Weidenschilling1977}, and once they cross the snowline, sublimation of their ice mantles can enhance the local gas-phase water abundance \citep{Ciesla2006}. High-resolution ALMA imaging has shown that gaps and rings are common in the millimeter (mm) dust distribution of disks \citep{Andrews2018, Long2018}, which may act as pressure bumps that trap dust pebbles and regulate this process. Disk models that track pebble drift together with ice sublimation and vapor transport predict that the strength and duration of the resulting water enrichment depend sensitively on whether and where such bumps intercept the drifting pebbles \citep{Booth2019, Kalyaan2021, Kalyaan2023, Mah2024}. Dust opacity, grain settling, and the broader thermal and chemical disk evolution may also modulate the observed water content \citep[e.g.,][]{Sellek2025, Houge2026, Calahan2026}.

The first observational evidence that outer disk structure shapes the inner water reservoir came from \citet{Banzatti2020}, who used Spitzer spectra to show that compact mm disks tend to have brighter water emission than their more extended and often structured counterparts. This built on earlier work linking the inner disk HCN-to-water emission ratio to outer disk dust mass \citep{Najita2013}.
JWST/MIRI has since enabled more direct tests. \citet{Banzatti2023} reported excess cool water emission near the water ice sublimation front ($\sim$200\,K) in two compact, mostly featureless disks relative to two large, gap-bearing ones, consistent with pebble drift proceeding largely unimpeded in the former (see also \citealt{Arulanantham2025, Romero-Mirza2024}). Larger samples have complicated this picture. Focusing on compact disks (mm disk size $<$ 40\,au) alone, \citet{Temmink2025} found cold water emission ranging from enhanced to depleted, indicating that outer disk size does not by itself set the inner disk water content. Similarly, \citet{Gasman2025} found that resolved mm gaps do not always suppress cold water emission across disks spanning a range of substructures. \citet{Banzatti2025} then pointed out a number of factors (including binarity, inner cavity, and cloud contamination) that may affect inner water emission properties beyond pebble drift. The observational link between outer disk dust dynamics and inner disk water content thus remains unsettled.

\citet{Krijt2025} moved further by adopting the radial location of the innermost dust gap as a measure of where drifting pebbles are finally intercepted (as suggested by models of \citealt{Kalyaan2023}).
They paired it with the flux ratio between cold and hot water emission lines ($F_{\rm cold}/F_{\rm hot}$, here cold and hot reflect different $E_{\rm up}$, as defined in \citealt{Banzatti2025}), which is assumed to measure the relative contribution of the cooler reservoir near the snowline and trace water delivered by inward-drifting icy pebbles. For a sample of 21 T~Tauri disks, they found that this ratio increases with the innermost gap location, as expected if a gap farther out encloses more interior pebble mass and sustains the pebble supply to the snowline for longer.

While the gap location serves as a local regulator to incoming pebbles, the gas-to-dust disk size ratio (usually $R_{\rm CO}/R_{\rm mm}$) offers a complementary global view, since drift preferentially depletes the mm dust disk while leaving the more extended CO gas disk largely intact. A high size ratio ($>4$) is widely taken as a signature of efficient inward pebble transport (e.g., \citealt{Facchini2019, Trapman2019}). This view is especially important for compact disks without strong substructures, where a small dust disk alone may not result from efficient drift, but could also reflect its initial condition \citep[e.g.,][]{Ricciardi2026}.

\begin{figure*}[th!]
    \centering
    \includegraphics[width=0.9\linewidth]{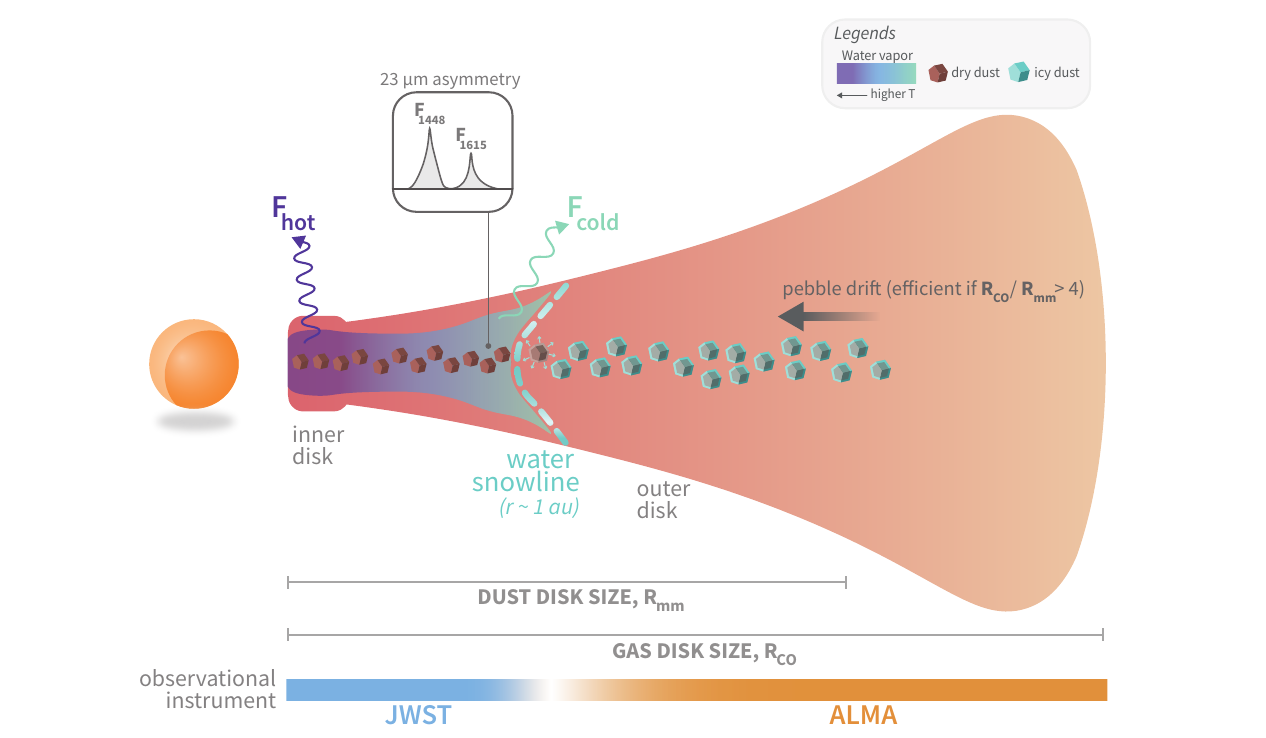}
\caption{Schematic illustration of key concepts and measurements used in this paper for a smooth disk around a Sun-like star. Gaps and rings are often found outside the water snowline in the outer disk and are not indicated. }
    \label{fig:Cartoon}
\end{figure*}

In this paper, we characterize dust substructures and $R_{\rm CO}/R_{\rm mm}$ in the outer disk for a sample of 51 disks drawn from four JWST/MIRI Cycle~1 programs, and connect these properties to inner disk water measurements.
With this large sample, we aim to examine whether and how the inner disk water reservoir responds to outer disk dust dynamics. Besides the $F_{\rm cold}/F_{\rm hot}$ diagnostic adopted in \citet{Krijt2025}, we also include the 23\,$\mu$m asymmetry, the flux ratio of two low-energy lines near 23.85\,$\mu$m whose similar $E_{\rm up}$ makes it a sensitive thermometer for gas near the snowline \citep{Banzatti2025}.
Figure~\ref{fig:Cartoon} outlines the key concepts and measurements used in this work. The paper is organized as follows. Section~\ref{sec:sample_obs} describes the sample and data. We then present the outer disk dust and gas properties in Section~\ref{sec:outer} and link these to the inner disk water diagnostics in Section~\ref{sec:link}. Section~\ref{sec:diss} discusses the results in the context of pebble drift, and Section~\ref{sec:summ} summarizes our findings.

\section{The Sample and Data} \label{sec:sample_obs}
Linking outer disk dust and gas distribution to inner disk water chemistry requires targets with both spatially resolved millimeter observations and mid-IR spectroscopy. We therefore start from the full set of JWST/MIRI Cycle 1 disk targets and identify which systems have the necessary mm continuum and CO observations to build the final sample. As described below, most targets are young ($<3$\,Myr) T Tauri stars with stellar masses of 0.3-0.8\,M$_{\odot}$.

\subsection{Sample description}
Our sample is drawn from four JWST Cycle 1 programs that requested MIRI/MRS observations for a total of 74 protoplanetary disks: GTO-1282 (PI: T.~Henning, \citealt{Henning2024}),  GO-1549 (PI: K.~Pontoppidan, \citealt{Pontoppidan2024}), GO-1584 (PI: C.~Salyk; co-PI: K.~Pontoppidan, \citealt{Arulanantham2025}), and GO-1640 (PI: A.~Banzatti, \citealt{Banzatti2023}).
We exclude disks around very low-mass stars (SpTy later than M4) for which outer disk properties are not readily accessible, and edge-on disks ($i>85\degr$) for which extracting inner chemical properties can be complicated by the disk geometry.
We also exclude FT\,Tau and IQ\,Tau, since their ALMA CO data only became publicly available recently, and IRS 48 because of its lopsided dust disk morphology. DF\,Tau, Sz\,50, and HD\,35929 are excluded due to the marginal detection of their CO emission (see Appendix~\ref{sec:appendix-uvfit} for their dust disk properties); DF\,Tau and Sz\,50 are also close binaries unresolved by JWST/MIRI (see \citealt{Grant2024_DFTau}). After applying these criteria, the final sample comprises 51 systems: 17 disks from Taurus, 11 from Ophiuchus, 10 from Lupus, 6 from Chamaeleon~I, 3 from Upper Sco, and 4 from various other regions.

\begin{figure*}[t!]
    \centering
    \includegraphics[width=\linewidth]{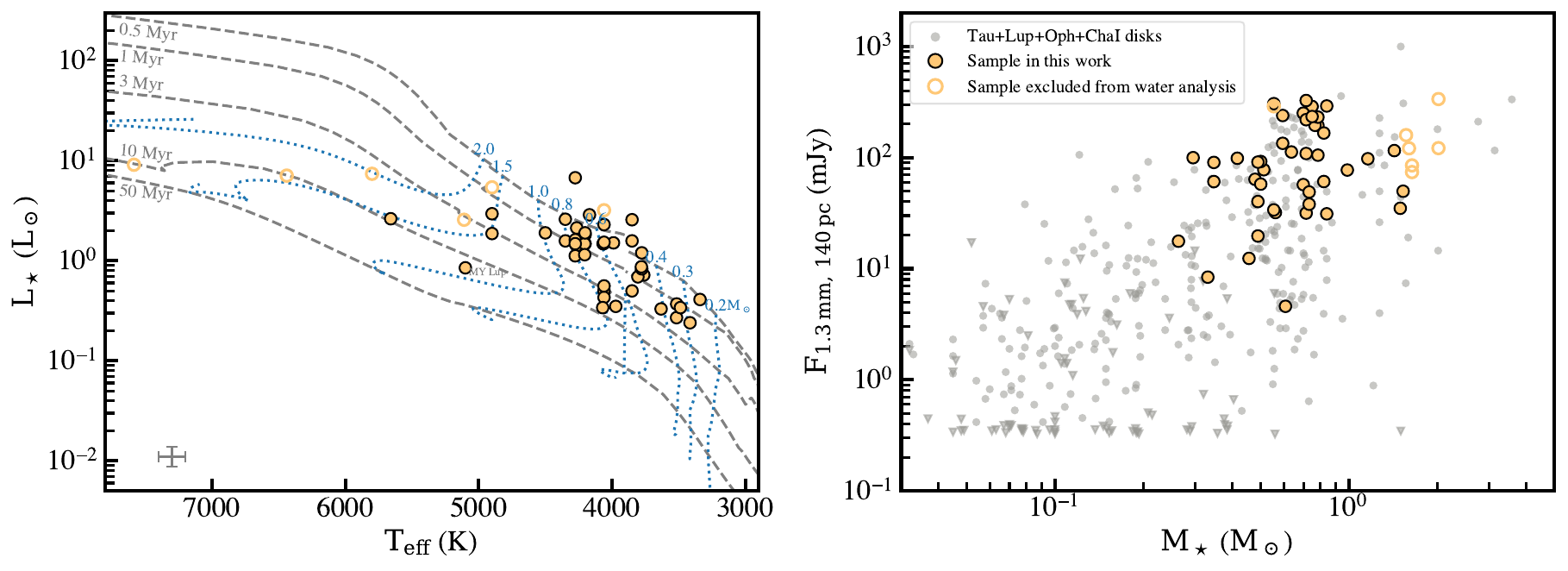}
    \caption{\textit{Left:} H-R diagram of our selected JWST/MIRI Cycle 1 sample, with the pre-main-sequence evolutionary tracks adopted from \citet{Baraffe2015} for $T_{\rm eff} < 3900$~K and \citet{Feiden2016} for hotter stars. This sample is dominated by T Tauri stars with ages younger than 3\,Myr. Open circles mark targets excluded entirely from the water analysis, for which both hot and cold water fluxes are unconstrained. MY Lup appears faint due to its inclined geometry. \textit{Right:} 1.3\,mm continuum flux (normalized to 140\,pc) vs. stellar mass for our sample compared to the young (1--3\,Myr) disk population compiled by \citet{Manara2023} (grey; downward triangles mark non-detections).}
    \label{fig:sample_hrd}
\end{figure*}

The sample spans stellar spectral types from A1 to M3.5, but is dominated by late K and early M stars (40/51 targets of K4-M3, with stellar masses of 0.3-0.8\,M$_{\odot}$). The left panel of Figure~\ref{fig:sample_hrd} shows the Hertzsprung-Russell diagram of our sample, which clearly demonstrates that this study mostly includes disks around young Sun-like stars with ages $<$3\,Myr. Individual stellar ages are derived using the \texttt{ysoisochrone} package \citep{Deng2025JOSS}, which incorporates the non-magnetic evolutionary models of \citet{Feiden2016} and \citet{Baraffe2015} for sources with hotter ($T_{\rm eff}>$3900\,K) and cooler stars, respectively.
Table~\ref{tab:source_prop} summarizes the stellar properties of the sample. Compared to the broader young disk population, we note that our sample is biased towards the brighter end of the outer disk mm flux distribution (right panel of Figure~\ref{fig:sample_hrd}), an intrinsic bias of sample selection in the early JWST cycle.

\begin{deluxetable*}{cccccccccccc}
\tabletypesize{\scriptsize}
\tablecaption{Host Stellar Properties\label{tab:source_prop}}
\tablewidth{0pt}
\tablehead{
\colhead{ID} & \colhead{Name} & \colhead{2MASS} & \colhead{Region} & \colhead{d} & \colhead{SpTy} & \colhead{$T_{\rm eff}$} & \colhead{$L_*$} & \colhead{$M_*$} & \colhead{Age} & \colhead{$\log(L_{\rm acc})$}  & \colhead{Refs} \\
\colhead{} & \colhead{} & \colhead{} & \colhead{} & \colhead{(pc)} & \colhead{} & \colhead{(K)} & \colhead{($L_\odot$)} & \colhead{($M_\odot$)} & \colhead{(Myr)} & \colhead{($L_\odot$)}  & \colhead{} \\
}
\startdata
1    & AA Tau       & J04345542$+$2428531        & Taurus     & 135      & M0.6  & 3762 & 0.72     & 0.49    & 0.8      & -0.85              &          \\
2    & AS 205 N     & J16113134$-$1838259        & UppSco     & 128      & K5    & 4265 & 2.13     & 0.87    & 0.6      & -0.07              & 1,2,3    \\
3    & AS 209       & J16491530$-$1422087        & Oph N 3a   & 121      & K5    & 4266 & 1.41     & 0.83    & 1.0      & -1.12              & 3,4,5    \\
4    & BP Tau       & J04191583$+$2906269        & Taurus     & 129      & M0.5  & 3777 & 0.83     & 0.48    & 0.7      & -1.17              & 3        \\
5    & CI Tau       & J04335200$+$2250301        & Taurus     & 160      & K5.5  & 4276 & 1.62     & 0.65    & 0.8      & 0.02               & 6        \\
6    & CX Tau       & J04144786$+$2648110        & Taurus     & 128      & M2.5  & 3487 & 0.34     & 0.33    & 1.1      & -2.56              & 3        \\
7    & CY Tau       & J04173372$+$2820468        & Taurus     & 126      & M2.5  & 3515 & 0.37     & 0.35    & 1.0      & -1.50              &          \\
8    & DG Tau       & J04270469$+$2606163        & Taurus     & 125      & K7    & 4060 & 3.2      & 0.71    & $<$0.5   & -0.15              &          \\
9    & DL Tau       & J04333906$+$2520382        & Taurus     & 159      & K5.5  & 4276 & 1.48     & 0.66    & 0.9      & -0.35              & 6        \\
10   & DM Tau       & J04334871$+$1810099        & Taurus     & 144      & M3    & 3415 & 0.24     & 0.29    & 1.4      & -1.92              & 3        \\
11   & DN Tau       & J04352737$+$2414589        & Taurus     & 128      & M0.3  & 3806 & 0.69     & 0.52    & 0.9      & -2.10              & 6        \\
12   & DoAr 25      & J16262367$-$2443138        & Oph L1688  & 139      & K6    & 4202 & 1.48     & 0.62    & 0.8      & \textless{}-2.13   &          \\
13   & DoAr 33      & J16273901$-$2358187        & Oph L1688  & 139      & K5    & 4350 & 1.58     & 0.69    & 1.0      & -1.77              &          \\
14   & DR Tau       & J04470620$+$1658428        & Taurus     & 192      & K6    & 4202 & 1.9      & 0.60    & 0.6      & 0.42               & 7        \\
15   & Elias 2-20   & J16261886$-$2428196        & Oph L1688  & 138      & M0    & 3850 & 2.57     & 0.59    & $<$0.5   & -0.09              &          \\
16   & Elias 2-24   & J16262407$-$2416134        & Oph L1688  & 136      & K5.5  & 4276 & 6.76     & 0.94    & $<$0.5   & 0.44               &          \\
17   & Elias 2-27   & J16264502$-$2423077        & Oph L1688  & 136      & K8    & 3989 & 1.51     & 0.56    & 0.6      & -0.57              &          \\
18   & FZ Tau       & J04323176$+$2420029        & Taurus     & 129      & M0.5  & 3777 & 1.2      & 0.51    & $<$0.5   & 0.34               & 8        \\
19   & GK Tau       & J04333456$+$2421058        & Taurus     & 129      & K6.5  & 4067 & 1.48     & 0.58    & 0.6      & -1.53              &          \\
20   & GO Tau       & J04430309$+$2520187        & Taurus     & 144      & M2.3  & 3516 & 0.27     & 0.35    & 1.6      & -2.00              & 9        \\
21   & GQ Lup       & J15491210$-$3539051        & Lupus      & 154      & K6    & 4205 & 1.48     & 0.61    & 0.8      & -0.36              & 3        \\
22   & GW Lup       & J15464473$-$3430354        & Lupus      & 155      & M1.5  & 3632 & 0.33     & 0.41    & 1.7      & -1.87              & 3        \\
23   & Haro 6-33    & J04413882$+$2556267        & Taurus     & 160      & M0.5  & 3850 & 0.5      & 0.64    & 1.8      & -1.30              & 10       \\
24   & HD 135344B   & J15154844$-$3709160        & UCL        & 135      & F4    & 6440 & 7.1      & 1.50    & 9.5      & -0.04              & 11,12    \\
25   & HD 142666    & J15564002$-$2201400        & UppSco     & 148      & A8    & 7586 & 9.12     & 1.58    & 9.3      & \textless{}-0.93   & 11       \\
26   & HD 143006    & J15583692$-$2257153        & UppSco     & 165      & K0    & 5110 & 2.57     & 1.48    & 3.0      & -0.50              &          \\
27   & HD 163296    & J17562128$-$2157218        & isolated   & 101      & A1    & 9333 & 16.98    & 2.04    & 10.7     & 0.08               & 11       \\
28   & HP Tau       & J04355277$+$2254231        & Taurus     & 171      & K4    & 4500 & 1.9      & 0.84    & 1.1      & -1.35              & 13       \\
29   & HT Lup       & J15451286$-$3417305        & Lupus      & 154      & K2    & 4900 & 5.42     & 1.32    & 0.8      & \ldots             &          \\
30   & IM Lup       & J15560921$-$3756057        & Lupus      & 158      & K5    & 4350 & 2.6      & 0.72    & 0.6      & -1.75              & 3        \\
31   & LkCa 15      & J04391779$+$2221034        & Taurus     & 156      & K5.5  & 4276 & 1.12     & 0.70    & 1.4      & -1.70              & 3        \\
32   & MY Lup       & J16004452$-$4155310        & Lupus      & 165      & K0    & 5100 & 0.85     & 1.19    & 2.0      & \textless{}-2.30   & 14       \\
33   & PDS 70       & J14081015$-$4123525        & LCC        & 113      & K7    & 3972 & 0.35     & 0.76    & 4.8      & -2.81              & 15,16    \\
34   & RNO 90       & J16340916$-$1548168        & $\rho$ Oph & 115      & G8    & 5660 & 2.63     & 1.68    & 9.3      & -0.19              & 3,5      \\
35   & RU Lup       & J15564230$-$3749154        & Lupus      & 159      & K7    & 4060 & 1.49     & 0.55    & 0.6      & -0.01              & 3        \\
36   & RW Aur       & J05074953$+$3024050        & Taurus     & 163      & K0    & 4900 & 2.95     & 1.48    & 1.6      & 0.39               & 6        \\
37   & RY Lup       & J15592838$-$4021513        & Lupus      & 158      & K2    & 4900 & 1.87     & 1.27    & 2.9      & -0.86              &          \\
38   & SR 21        & J16271027$-$2419127        & $\rho$ Oph & 138      & F7    & 5800 & 7.4      & 1.67    & 4.3      & -0.70              & 12,17,18 \\
39   & SR 4         & J16255615$-$2420481        & Oph L1688  & 134      & K6    & 4202 & 1.82     & 0.61    & 0.6      & -0.12              &          \\
40   & SY Cha       & J10563044$-$7711393        & ChaI       & 182      & K7    & 4060 & 0.56     & 0.67    & 2.6      & -2.13              &          \\
41   & Sz 129       & J15591647$-$4157102        & Lupus      & 161      & K7    & 4060 & 0.43     & 0.73    & 4.3      & -1.14              &          \\
42   & Sz 98        & J16082249$-$3904464        & Lupus      & 156      & K7    & 4060 & 1.53     & 0.55    & 0.6      & -1.53              & 3        \\
43   & TW Cha       & J10590108$-$7722407        & ChaI       & 182      & K7    & 4060 & 0.5      & 0.70    & 3.3      & -1.54              &          \\
44   & TW Hya       & J11015191$-$3442170        & TWA        & 60       & M0.5  & 4070 & 0.34     & 0.81    & 6.4      & -1.53              & 3,12     \\
45   & V1094 Sco    & J16083617$-$3923024        & Lupus      & 156      & K6    & 4205 & 1.15     & 0.64    & 1.1      & -1.02              &          \\
46   & VW Cha       & J11080148$-$7742288        & ChaI       & 191      & K7    & 4060 & 2.3      & 0.75    & $<$0.5   & 0.13               & 19       \\
47   & VZ Cha       & J11092379$-$7623207        & ChaI       & 191      & M0.5  & 3780 & 0.79     & 0.50    & 0.7      & -0.33              &          \\
48   & WaOph 6      & J16484562$-$1416359        & Oph N 3a   & 123      & K6    & 4169 & 2.88     & 0.68    & $<$0.5   & -0.66              & 3,4      \\
49   & WSB 52       & J16273942$-$2439155        & Oph L1688  & 136      & M0    & 3850 & 1.58     & 0.55    & $<$0.5   & -1.11              &          \\
50   & WX Cha       & J11095873$-$7737088        & ChaI       & 191      & M0.5  & 3780 & 0.87     & 0.47    & 0.6      & 0.35               & 20       \\
51   & XX Cha       & J11113965$-$7620152        & ChaI       & 192      & M3.5  & 3340 & 0.41     & 0.25    & 0.7      & -2.34              & 20       \\
\enddata
\tablecomments{Distances are calculated from inverting Gaia DR3 parallaxes \citep{GaiaCollaboration2023}. Stellar properties are taken from the compiled table in \citet{Manara2023} when available, with masses based on the same tracks as our ages; otherwise see references below. The MY Lup disk is very inclined thus it would appear fainter and older, for which we assign an average age of 2\,Myr. UCL = Upper Centaurus--Lupus, LCC = Lower Centaurus--Crux, TWA = TW Hydrae Association.}
\tablerefs{1=\citet{Luhman2012}, 2=\citet{Barenfeld2016}, 3=\citet{Fang2018}, 4=\citet{Andrews2009}, 5=\citet{Salyk2013}, 6=\citet{Gangi2022}, 7=\citet{Banzatti2014}, 8=\citet{McClure2019}, 9=\citet{Simon2016}, 10=\citet{White2004}, 11=\citet{Fairlamb2015}, 12=\citet{HH2014}, 13=\citet{Valenti1993}, 14=\citet{Alcala2017}, 15=\citet{PecautMamajek2016}, 16=\citet{Skinner2022}, 17=\citet{Yang2023}, 18=\citet{Manara2014}, 19=\citet{Zsidi2022}, 20=\citet{Fiorellino2025}}
\end{deluxetable*}

\subsection{Data for outer disk properties}
To characterize the outer disk properties relevant to this study, we make use of new and archival ALMA/SMA datasets that provide high-resolution continuum and $^{12}$CO ($J=2-1$ or $J=3-2$) line observations. These data enable us to assess disk dust substructures and gas-to-dust size ratios with sufficient spatial resolution and sensitivity.

\subsubsection{New ALMA observations} \label{sec:new_data}
The ALMA program 2021.1.00854.S (PI: F.~Long) targeted 11 disks (FZ Tau, TW Cha, VZ Cha, Haro 6-33, CY Tau, DF Tau, Sz 50, VW Cha, WX Cha, XX Cha, and HD 35929) that would have mid-IR spectra from approved JWST/MIRI Cycle 1 programs but lacked high-resolution mm images. Observations were carried out using ALMA Band 6 receivers \citep{Kerr2004_b6} to capture dust continuum emission in two spectral windows at 218 and 233\,GHz, along with $^{12}$CO $J=2-1$ line.
This program was designed to use both long-baseline (C-8, $\sim$25\,min on-source) and short-baseline (C-5, $\sim$5\,min) configurations to obtain high-fidelity images with $\sim0\farcs05$ resolution. Short-baseline data, covering baselines of 15$\sim$1300\,m, were taken during two sessions in December 2021 and July 2022, and are available for all targets. Long-baseline data, collected between October 14 and 29, 2021 (covering baselines of 92$\sim$8500\,m) show more diversities in execution: FZ Tau was observed three times, yielding a total on-source time of 87\,min; the execution block (EB) including TW Cha, VW Cha, and WX Cha was only semi-passed in QA2 with an on-source time of 13\,min for each; VZ Cha and XX Cha each have one successful EB with  13\,min of integration time (an additional semi-passed EB was excluded); CY Tau was not observed, and archival data with lower spatial resolution were used instead; Sz 50 was semi-passed with 25\,min of integration time; the remaining targets were observed as planned. As a result, the final resolution and sensitivity vary across the sample. Details are summarized in Table~\ref{tab:data_info} in the Appendix.

The $^{12}$CO $J=2-1$ data for BP\,Tau and HP\,Tau were obtained from the ALMA program 2021.1.01050.S (PI: F.~Long), with baseline coverage of 15-3637\,m and 15-2517m, respectively. Both targets were observed twice, yielding total on-source times of 46\,min for BP\,Tau and 21\,min for HP\,Tau. For the mm continuum image of BP\,Tau, data from the continuum spectral window in the program above were combined with high-resolution data from ALMA program 2019.1.00607.S (PI: F.~Long), which covered baselines of 178-16196\,m.
For RNO\,90, Band 8 observations from ALMA program 2015.1.00847.S (PI: F.~Du) provide the highest available spatial resolution ($\sim0\farcs1$), from which the two continuum spectral windows at a mean frequency of 402\,GHz were adopted for the mm continuum analysis. The CO data for RNO\,90 were obtained separately from ALMA program 2015.1.00168.S (PI: G.~Blake).

For each EB, initial data reduction was performed using the scripts/pipelines provided by the observatory with the specified CASA versions \citep{CASATeam2022PASP}, to carry out standard bandpass, gain, and flux calibrations. Subsequent self-calibration and imaging were conducted using CASA 6.5, basically following the procedure outlined in \citet{Andrews2018}.
First, channels containing strong line emission were flagged, and the remaining line-free channels were averaged to generate the pseudo-continuum visibilities.
For disks with multiple EBs, we created an initial continuum image for each EB with the \texttt{tclean} task and determined the disk emission center with 2D Gaussian fitting. Phase alignment to a common center was achieved using the \texttt{phaseshift} and \texttt{fixplanets} tasks. To ensure consistent amplitude scaling across EBs, the \texttt{gencal} task was used to rescale the continuum visibilities to match the flux of a chosen reference EB, for which
the short-baseline EB was typically used, since some disks had only semi-passed long-baseline data\footnote{The scaling factor in \texttt{gencal} for semi-passed EBs can be as low as 0.7--0.8, which could lead to an overestimated noise level in the combined image. However, the disk morphology should remain unaffected.}.

Self-calibration for each disk began with the (phase-shifted and flux-aligned) short-baseline EBs, applying multiple rounds of phase self-calibration until no further improvement in the image peak signal-to-noise ratio was observed. One round of amplitude self-calibration was attempted, but it typically did not yield significant improvement. Then we conducted a combined self-calibration for both short- and long-baseline datasets. For disks with semi-passed data, self-calibration was carried out on individual EBs prior to phase alignment and flux rescaling. Finally, the self-calibration solutions derived from the averaged continuum visibilities were applied to the original full measurement sets, which include line emission. Final images were generated from the self-calibrated visibilities using auto-masking,
and customized weighting schemes for individual disks. Continuum subtraction for the $^{12}$CO line was performed using the \texttt{uvcontsub} task before imaging.

\subsubsection{Archival Data} \label{sec:archival_data}
Archival data were used for the remaining targets. Many of these disks are part of the DSHARP and exoALMA Large Programs, from which we directly adopted the publicly released, self-calibrated measurement sets and images \citep{Andrews2018, Teague2025}. For TW~Hya and RW~Aur, we used the reduced data products presented by \citet{Huang2018_twhya} and \citet{Kurtovic2024}, respectively. Continuum and CO line (from SMA observations) data for GO~Tau and DL~Tau were taken from \citet{Long2018} and \citet{Long2022gas}; continuum data for a few other Taurus disks were also taken directly from \citet{Long2018} and \citet{Long2019}.
For all other disks, we retrieved and reprocessed the highest-quality datasets available in the ALMA archive, applying self-calibration and generating images following the procedure outlined above. We used the $J=2-1$ transition of CO for all disks, except for seven disks for which only $J=3-2$ data are available or readily accessible: five from exoALMA, TW~Hya, and RNO~90. Models show that the gas outer radii derived from the two transitions should differ by less than 10\% \citep{Trapman2019}.
A summary of the datasets and properties of their image products is provided in Table~\ref{tab:data_info} of the Appendix.

\subsection{JWST data for inner disk chemistry}
The details of the JWST observation setups can be found in earlier published works (e.g., \citealt{Arulanantham2025, Banzatti2025, Gasman2025, Temmink2025}).
All MIRI/MRS data used in this work were reduced with the JDISCS pipeline version 9.1, following the procedure described in \citet{Pontoppidan2024}, \citet{Banzatti2025} and \citet{Salyk2026}. The pipeline follows the standard JWST MIRI/MRS calibration pipeline through stage 2b, then uses asteroid spectra observed as calibrators to provide high-quality fringe removal.

Prior to line analysis, continuum subtraction was performed using median smoothing combined with a second-order Savitzky--Golay filter following \citet{Pontoppidan2024}, with a final offset correction based on line-free regions applied to each spectrum (see details in \citealt{Banzatti2025}). Line fluxes are then measured using the \texttt{iSLAT} software \citep{Jellison2024}.  The two main water emission diagnostics used in this work are the cold-to-hot flux ratio $F_{\rm cold}/F_{\rm hot}$ and the 23\,$\mu$m line asymmetry $F_{1448}/F_{1615}$. The hot water component is measured from the isolated line at 17.324\,$\mu$m with $E_{\rm up} \approx 6000$\,K, and the cold water flux is taken as the sum of two H$_2$O transitions near 23.85\,$\mu$m with $E_{\rm up}$ of 1448 and 1615\,K \citep{Banzatti2025}. The 23\,$\mu$m line asymmetry denotes the ratio of the two cold water lines. All water line properties are taken from the HITRAN database \citep{Gordon2022}.

\begin{figure*}[th!]
    \centering
    \includegraphics[width=\linewidth]{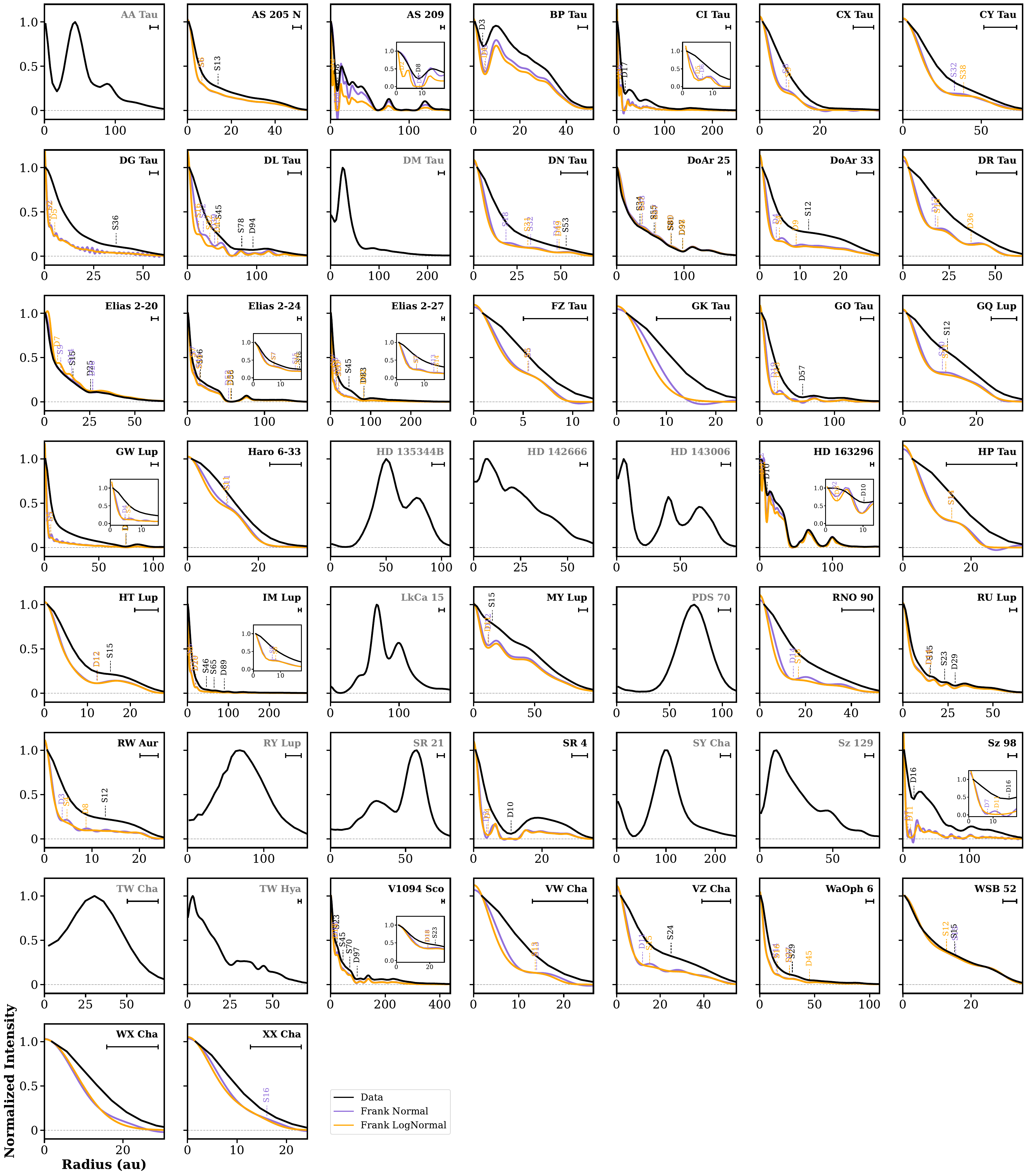}
    \caption{Normalized mm continuum radial intensity profiles for all 51 disks. Black curves are the deprojected, azimuthally averaged image-plane profiles. For full disks, purple/orange curves show the \texttt{frank} normal/log-normal visibility-plane fits, with dashed vertical lines marking the innermost gap (`D', radius in au) and any interior shoulders (`S') in matching colors; fits are omitted for transition disks (grey names) for clarity. The upper-right bar in each panel gives the average CLEAN beam resolution. Zoomed-in insets show the innermost regions of a few extended disks.}
    \label{fig:cont_rp}
\end{figure*}

\section{Outer Disk Properties} \label{sec:outer}

\begin{deluxetable*}{cccccccccccc}
\tabletypesize{\scriptsize}
\tablecaption{Outer Disk Properties from ALMA\label{tab:alma_disk}}
\tablewidth{0pt}
\tablehead{
\colhead{Name}
  & \colhead{$i$} & \colhead{PA}
  & \colhead{$R_{\rm mm}^{\rm img}$} & \colhead{$R_{\rm mm}^{\rm frank}$}
  & \colhead{$R_{\rm CO}^{\rm img}$} & \colhead{$R_{\rm CO}^{\rm deconv}$}
  & \colhead{$\mathcal{R}_{\rm img}$} & \colhead{$\mathcal{R}_{\rm dec}$}
  & \colhead{CO morph.}
  & \multicolumn{2}{c}{Innermost feature} \\
\colhead{}
  & \colhead{(deg)} & \colhead{(deg)}
  & \colhead{(au)} & \colhead{(au)}
  & \colhead{(au)} & \colhead{(au)}
  & \colhead{} & \colhead{}
  & \colhead{}
  & \colhead{type} & \colhead{location (au)}
}
\startdata
AA Tau & $58.7$ & $92.7$ & $140.7\pm5.8$ & $136.9$ & $456.6\pm10.2$ & $456.1\pm1.1$ & $3.2\pm0.1$ & $3.3\pm0.1$ & \nodata & TD & $17.6$ \\
AS 205 N & $20.1$ & $114.0$ & $45.6\pm2.0$ & $44.7$ & \nodata & \nodata & \nodata & \nodata & binary & S & $6.6$ \\
AS 209 & $35.0$ & $85.8$ & $127.2\pm2.2$ & $127.7$ & $275.9\pm5.1$ & $275.7\pm0.6$ & $2.2\pm0.1$ & $2.2$ & \nodata & D & $8.7\pm0.4$ \\
BP Tau & $38.2$ & $151.1$ & $43.1\pm2.0$ & $41.4$ & $112.9\pm12.4$ & $110.2\pm2.1$ & $2.6\pm0.3$ & $2.7\pm0.3$ & \nodata & D & $4.9\pm0.1$ \\
CI Tau & $50.0$ & $11.2$ & $208.5\pm4.7$ & $184.3$ & $416.8\pm35.3$ & $410.9\pm5.9$ & $2.0\pm0.2$ & $2.2\pm0.2$ & \nodata & D & $5.9\pm0.7$ \\
CX Tau & $55.1$ & $66.2$ & $33.1\pm3.5$ & $25.7$ & $107.9\pm8.7$ & $106.5\pm1.1$ & $3.2\pm0.4$ & $4.1\pm0.3$ & \nodata & S & $9.2\pm0.4$ \\
CY Tau & $26.1$ & $153.1$ & $63.6\pm10.5$ & $60.5$ & $220.6\pm14.6$ & $218.8\pm3.3$ & $3.5\pm0.6$ & $3.6\pm0.2$ & \nodata & S & $35.9\pm3.0$ \\
DG Tau & $37.3$ & $135.0$ & $50.6\pm2.2$ & $50.2$ & \nodata & \nodata & \nodata & \nodata & cloud & S & $2.8\pm0.2$ \\
DL Tau & $45.0$ & $52.1$ & $144.2\pm9.8$ & $146.8$ & $588.6\pm91.2$ & $560.5\pm15.7$ & $4.1\pm0.7$ & $3.8\pm0.6$ & \nodata & S & $19.4\pm3.4$ \\
DM Tau & $35.9$ & $155.6$ & $205.9\pm6.2$ & $182.2$ & $879.0\pm10.8$ & $878.7\pm0.7$ & $4.3\pm0.1$ & $4.8\pm0.1$ & \nodata & TD & $8.3$ \\
DN Tau & $35.2$ & $79.2$ & $57.3\pm7.7$ & $55.9$ & $180.1\pm13.1$ & $178.3\pm2.6$ & $3.1\pm0.5$ & $3.2\pm0.2$ & \nodata & S & $31.9\pm0.8$ \\
DoAr 25 & $67.4$ & $110.6$ & $148.4\pm2.1$ & $148.9$ & $359.7\pm6.3$ & $359.5\pm1.4$ & $2.4\pm0.1$ & $2.4$ & \nodata & S & $37.8\pm0.5$ \\
DoAr 33 & $41.8$ & $81.1$ & $24.7\pm2.1$ & $24.5$ & $64.2\pm6.3$ & $63.0\pm0.6$ & $2.6\pm0.3$ & $2.6\pm0.3$ & \nodata & S & $4.5\pm0.4$ \\
DR Tau & $5.4$ & $3.4$ & $54.0\pm10.8$ & $48.2$ & \nodata & \nodata & \nodata & \nodata & spiral & S & $18.2\pm0.7$ \\
Elias 2-20 & $49.0$ & $153.2$ & $55.2\pm1.9$ & $53.8$ & \nodata & \nodata & \nodata & \nodata & outflow & D & $8.2\pm0.9$ \\
Elias 2-24 & $29.0$ & $45.7$ & $129.6\pm2.4$ & $128.0$ & \nodata & \nodata & \nodata & \nodata & cloud & S & $7.5$ \\
Elias 2-27 & $56.2$ & $118.8$ & $248.7\pm3.4$ & $230.5$ & \nodata & \nodata & \nodata & \nodata & cloud & S & $7.1$ \\
FZ Tau & $13.4$ & $169.3$ & $10.1\pm3.2$ & $7.6$ & \nodata & \nodata & \nodata & \nodata & spiral & S & $5.5$ \\
GK Tau & $40.2$ & $119.9$ & $20.1\pm7.5$ & $11.0$ & $70.0\pm13.6$ & $64.6\pm1.4$ & $3.5\pm1.6$ & $5.9\pm1.2$ & \nodata & \nodata & \nodata \\
GO Tau & $53.9$ & $20.9$ & $133.8\pm8.8$ & $142.4$ & $1012.6\pm82.6$ & $999.3\pm12.2$ & $7.6\pm0.8$ & $7.0\pm0.6$ & \nodata & D & $21.6\pm2.2$ \\
GQ Lup & $58.0$ & $169.0$ & $27.1\pm3.5$ & $22.9$ & $225.2\pm14.9$ & $223.3\pm3.1$ & $8.3\pm1.2$ & $9.7\pm0.7$ & \nodata & S & $11.2\pm0.5$ \\
GW Lup & $38.7$ & $37.6$ & $91.6\pm3.4$ & $90.4$ & $267.2\pm7.7$ & $266.8\pm2.3$ & $2.9\pm0.1$ & $3.0\pm0.1$ & \nodata & S & $5.3\pm0.6$ \\
Haro 6-33 & $60.1$ & $161.0$ & $28.2\pm4.5$ & $21.6$ & \nodata & \nodata & \nodata & \nodata & outflow & S & $11.2\pm0.2$ \\
HD 135344B & $16.1$ & $62.9$ & $89.9\pm5.6$ & $89.1$ & $202.5\pm10.1$ & $201.5\pm0.4$ & $2.2\pm0.2$ & $2.3\pm0.1$ & \nodata & TD & $13.6$ \\
HD 142666 & $62.2$ & $162.1$ & $53.1\pm2.0$ & \nodata & $170.9\pm5.2$ & $170.6\pm0.7$ & $3.2\pm0.1$ & \nodata & \nodata & TD & $3.7$ \\
HD 143006 & $18.6$ & $169.0$ & $78.4\pm3.7$ & $77.2$ & $153.6\pm4.8$ & $153.3\pm1.0$ & $2.0\pm0.1$ & $2.0\pm0.1$ & \nodata & TD & $2.7$ \\
HD 163296 & $46.7$ & $133.3$ & $137.3\pm2.2$ & $137.5$ & $477.7\pm5.0$ & $477.6\pm0.3$ & $3.5\pm0.1$ & $3.5$ & \nodata & D & $2.8\pm0.3$ \\
HP Tau & $18.3$ & $56.6$ & $29.2\pm10.3$ & $19.7$ & \nodata & \nodata & \nodata & \nodata & outflow & S & $14.3$ \\
HT Lup & $48.1$ & $166.1$ & $23.2\pm2.7$ & $21.7$ & $71.9\pm4.0$ & $71.5\pm0.6$ & $3.1\pm0.4$ & $3.3\pm0.2$ & \nodata & D & $12.3\pm0.1$ \\
IM Lup & $47.5$ & $144.5$ & $243.8\pm3.5$ & $246.8$ & $803.4\pm9.5$ & $803.2\pm1.7$ & $3.3\pm0.1$ & $3.2$ & \nodata & S & $7.6\pm0.6$ \\
LkCa 15 & $50.4$ & $61.9$ & $145.7\pm3.9$ & $145.9$ & $685.0\pm11.7$ & $684.6\pm1.1$ & $4.7\pm0.1$ & $4.7\pm0.1$ & \nodata & TD & $14.6$ \\
MY Lup & $73.2$ & $58.8$ & $81.7\pm3.6$ & $79.7$ & $174.9\pm7.5$ & $174.3\pm1.0$ & $2.1\pm0.1$ & $2.2\pm0.1$ & \nodata & D & $12.1\pm0.3$ \\
PDS 70 & $51.7$ & $160.4$ & $94.6\pm5.8$ & $91.4$ & $216.4\pm7.6$ & $215.8\pm0.6$ & $2.3\pm0.2$ & $2.4\pm0.1$ & \nodata & TD & $21.3$ \\
RNO 90 & $51.7$ & $172.6$ & $43.6\pm6.9$ & $47.4$ & $125.1\pm16.8$ & $120.6\pm2.0$ & $2.9\pm0.6$ & $2.5\pm0.4$ & \nodata & S & $15.8\pm1.1$ \\
RU Lup & $18.8$ & $121.0$ & $55.3\pm2.0$ & $55.5$ & \nodata & \nodata & \nodata & \nodata & spiral & D & $14.6$ \\
RW Aur & $54.9$ & $39.4$ & $21.0\pm1.9$ & $20.5$ & \nodata & \nodata & \nodata & \nodata & binary & S & $4.2\pm0.5$ \\
RY Lup & $68.0$ & $109.0$ & $130.8\pm10.2$ & $127.1$ & $277.9\pm14.3$ & $276.4\pm1.7$ & $2.1\pm0.2$ & $2.2\pm0.1$ & \nodata & TD & $34.3$ \\
SR 21 & $15.0$ & $16.0$ & $66.5\pm2.4$ & $65.3$ & $94.8\pm7.2$ & $93.7\pm1.1$ & $1.4\pm0.1$ & $1.4\pm0.1$ & \nodata & TD & $15.2$ \\
SR 4 & $22.0$ & $18.0$ & $29.2\pm2.3$ & $28.5$ & $81.9\pm6.6$ & $80.8\pm0.9$ & $2.8\pm0.3$ & $2.8\pm0.2$ & \nodata & D & $4.0\pm0.2$ \\
SY Cha & $50.7$ & $165.6$ & $202.6\pm10.3$ & $197.7$ & $514.7\pm13.6$ & $514.0\pm0.9$ & $2.5\pm0.1$ & $2.6\pm0.1$ & \nodata & TD & $32.4$ \\
Sz 129 & $34.1$ & $151.0$ & $68.4\pm3.0$ & $65.8$ & $132.8\pm7.8$ & $131.9\pm1.0$ & $1.9\pm0.1$ & $2.0\pm0.1$ & \nodata & TD & $5.7$ \\
Sz 98 & $47.0$ & $110.0$ & $149.6\pm6.5$ & $138.4$ & $295.9\pm14.5$ & $294.5\pm1.4$ & $2.0\pm0.1$ & $2.1\pm0.1$ & \nodata & D & $9.7\pm2.0$ \\
TW Cha & $29.6$ & $129.9$ & $60.6\pm9.4$ & $53.1$ & $209.3\pm21.6$ & $205.4\pm7.6$ & $3.5\pm0.7$ & $3.9\pm0.4$ & \nodata & TD & $15.2$ \\
TW Hya & $5.0$ & $152.0$ & $58.6\pm0.9$ & \nodata & $184.3\pm4.0$ & $184.1\pm0.1$ & $3.1\pm0.1$ & \nodata & \nodata & TD & $1.5$ \\
V1094 Sco & $53.0$ & $109.0$ & $365.4\pm4.6$ & $379.9$ & $967.7\pm12.9$ & $967.4\pm3.1$ & $2.6\pm0.1$ & $2.5$ & \nodata & D & $18.8$ \\
VW Cha & $26.6$ & $36.0$ & $22.2\pm6.1$ & $17.4$ & \nodata & \nodata & \nodata & \nodata & binary & S & $13.8\pm0.2$ \\
VZ Cha & $18.8$ & $31.3$ & $45.8\pm6.5$ & $43.5$ & \nodata & \nodata & \nodata & \nodata & outflow & S & $13.5\pm1.7$ \\
WaOph 6 & $47.3$ & $174.2$ & $90.9\pm3.4$ & $90.9$ & $297.3\pm7.5$ & $296.9\pm0.7$ & $3.3\pm0.1$ & $3.3\pm0.1$ & \nodata & S & $15.6\pm0.6$ \\
WSB 52 & $54.4$ & $138.4$ & $29.2\pm2.0$ & $28.6$ & \nodata & \nodata & \nodata & \nodata & outflow & S & $14.1\pm1.4$ \\
WX Cha & $53.7$ & $155.5$ & $25.4\pm6.6$ & $18.5$ & $77.5\pm15.5$ & $71.2\pm2.1$ & $3.0\pm1.1$ & $3.8\pm0.8$ & \nodata & \nodata & \nodata \\
XX Cha & $27.0$ & $138.6$ & $20.2\pm5.2$ & $16.7$ & $82.8\pm12.1$ & $79.2\pm1.3$ & $4.1\pm1.3$ & $4.7\pm0.7$ & \nodata & \nodata & \nodata \\
\enddata
\tablecomments{Sources are ordered following Table~\ref{tab:source_prop}. $\mathcal{R}_{\rm img} = R_{\rm CO}^{\rm img}/R_{\rm mm}^{\rm img}$; $\mathcal{R}_{\rm dec} = R_{\rm CO}^{\rm deconv}/R_{\rm mm}^{\rm frank}$. CO morphology is blank for Keplerian-dominated disks; otherwise, the dominant non-Keplerian feature is noted. Inner feature types include $S$~=~inflection/shoulder, $D$~=~dark gap, TD~=~transition disk. \texttt{Frank} fitting results were not shown for HD 142666 and TW Hya as the fits do not converge well. Since both disks are transition disks, the lack of \texttt{Frank} results does not affect our analysis.}
\end{deluxetable*}

\subsection{Disk Morphology}
\subsubsection{Dust Disks}
Figure~\ref{fig:cont_rp} shows the deprojected, azimuthally averaged radial profiles for the mm dust continuum emission of our sample.
Disk inclination and position angles used for deprojection are adopted from the literature when available (see Table~\ref{tab:alma_disk}). For disks with new ALMA observations as described in Section 2.2.1, such geometric information is often unavailable in the literature due to the lack of previous high-resolution imaging. In these cases, we instead derive it from Nuker profile fitting in the visibility plane, as described in Appendix~\ref{sec:appendix-uvfit}.

Two types of substructures can be identified from the radial profiles: gaps/rings, defined as local minima/maxima; shoulders (in the absence of a gap-ring pair), defined as local maxima in the gradient of the radial profile and appearing as inflection-point-like features (see the illustrative definitions in \citealt{Huang2024}). We identify 13 transition disks where the ring peak flux exceeds that at the disk center: HD~142666, TW~Hya, TW~Cha, SR~21, Sz~129, HD~143006, HD~135344B, PDS~70, RY~Lup, AA~Tau, LkCa~15, SY~Cha, DM~Tau. An increasing number of transition disks reveal an inner disk component when imaged at high resolution and sensitivity. This inner disk may include both thermal dust emission and free-free emission \citep{Rota2024}.
There are 11 disks that display only shoulder features.
In addition, 11 disks appear smooth at the current angular resolution, with disk radii only $\sim$2--3 beam widths on average.

To better identify and characterize disk substructures, we used the super-resolution tool \texttt{frank}, which fits non-parametric 1D models directly in the visibility domain \citep{Jennings2020}. We set $R_{max}=2 \times R_{mm}^{img}$, where $R_{mm}^{img}$ is the disk radius derived from the image plane (see section~\ref{sec:disk size}) to encompass all possible emission, and adopted conservative hyperparameters to minimize fitting artifacts: $\alpha=1.3$ and $\omega_{smooth}=0.01$. Here, $\alpha$ sets the signal-to-noise ratio threshold beyond which the model no longer fits the data, and $\omega_{smooth}$ controls the level of smoothing applied to damp oscillations.
To test the robustness of detected features, we performed \texttt{frank} fitting in both normal and logarithmic (non-negative) intensity space.

Figure~\ref{fig:cont_rp} shows the \texttt{frank} fitting results, which reveal more and sharper features than seen in the image plane. For clarity, we omit \texttt{frank} fits for transition disks, since we do not attempt to identify additional substructure in these disks.
The non-negative fits generally produce shallower profiles with fewer oscillations. Theoretical models that explore how dust traps regulate the inner disk water reservoir suggest that the innermost features are most important \citep[e.g.,][]{Kalyaan2023, Krijt2025}, which are thus highlighted in Figure~\ref{fig:cont_rp} for individual disks and summarized in Table~\ref{tab:alma_disk}. In many disks, the innermost features are located at smaller radii in the \texttt{frank} profiles than in the image. We consider a feature reliable only if it appears in both \texttt{frank} fits. If the two fits give slightly different locations, with differences smaller than 0$\farcs$05 ($\sim$5--10\,au for our sample), we treat them as the same feature and use the location difference as a measure of the uncertainty; the majority of differences are within 3\,au. If one fit shows a shoulder-like emission while the other shows a gap, we adopt the more conservative interpretation and classify the feature as a shoulder.
Among the 11 disks that appear featureless in the image plane, three remain featureless with \texttt{frank} (GK Tau, XX Cha, WX Cha) at angular resolutions of 0.$''$05-0.$''$1, and the rest now show shoulder-like emission (with an additional outer gap-ring pair for DR Tau).

\begin{figure*}[!t]
  \includegraphics[width=\linewidth, trim=0 100 100 100, clip]{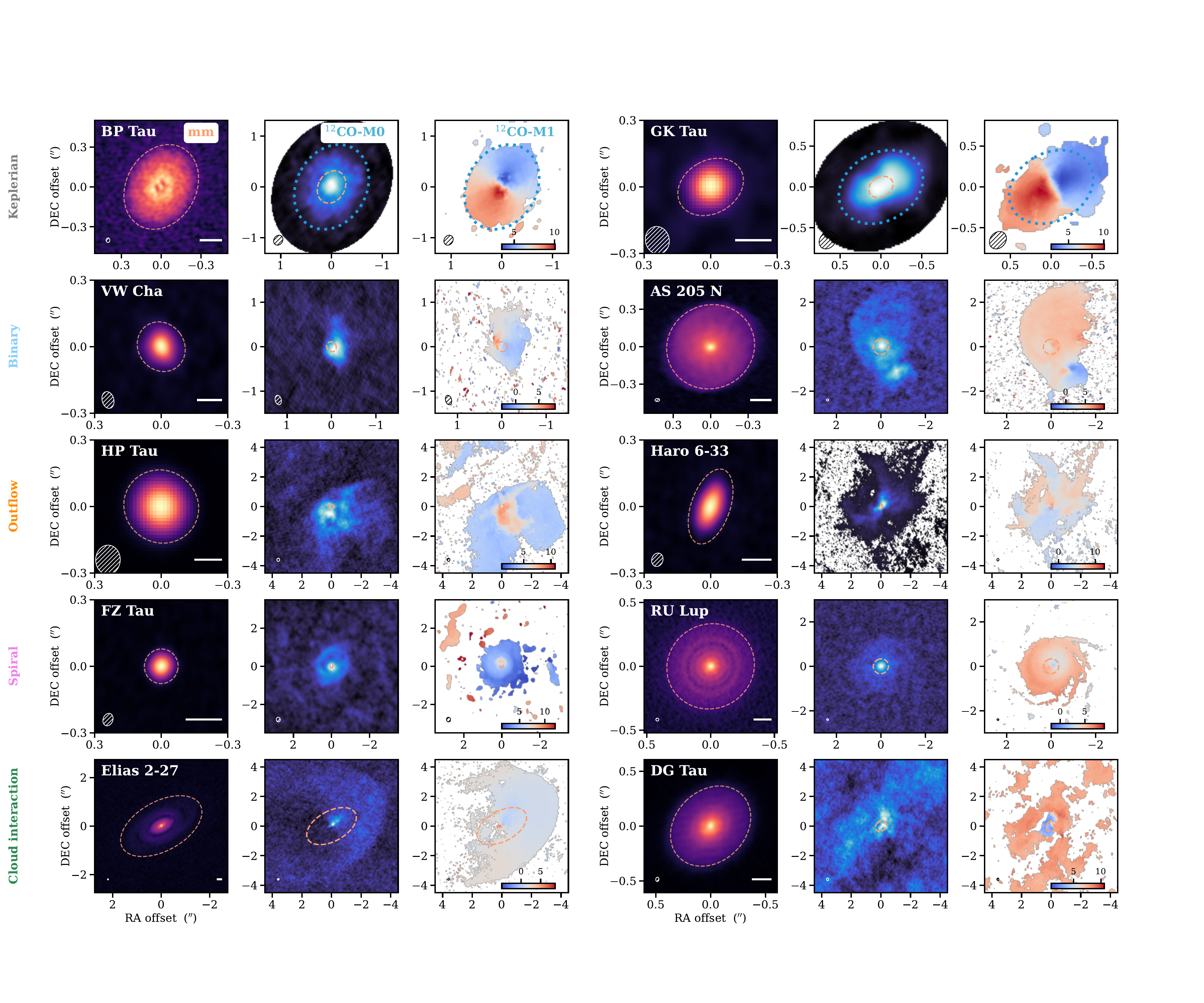}
  \caption{Examples of ALMA mm continuum and CO gas maps, with each row for one type of CO gas morphology. For each source, three panels are included: (\textit{left}) mm continuum image; (\textit{middle}) CO moment 0 map; (\textit{right}) CO moment 1 map. Beam size is indicated in the lower corner of each panel, and the white bar in the continuum panel indicates a scale of 20\,au. The orange and blue dashed ellipses mark $R_{\rm mm}$ and $R_{\rm CO}$ (see definition in Sec.~\ref{sec:disk size}), respectively.}
  \label{fig:cont-co-panels-example}
\end{figure*}

\subsubsection{CO Gas Disks} \label{sec:co-disk}
Among the 51 disks with detected CO gas emission, 14 show prominent non-Keplerian kinematic features upon careful visual inspection of the CO image cubes.
To characterize the gas emission morphology, we produced moment 0 (velocity-integrated intensity) and moment 1 (velocity) maps for all sources using the scripts provided by the AGEPRO collaboration
\citep{Trapman2025, Trapman2025zenodo}, which build on the \texttt{bettermoments} Python package \citep{bettermoments2018zndo}. A subset of maps is displayed in Figure~\ref{fig:cont-co-panels-example} as examples and the rest are shown in Figure~\ref{fig:cont-co-panels-normal-remaining} in the Appendix~\ref{sec:appendix-data}.
The masking approach varies with the CO emission properties.
For the 37 Keplerian-dominated disks with characteristic butterfly pattern of blue- and red-shifted emission in the velocity map, we constructed a Keplerian mask using the \texttt{keplerian$\_$mask} function from the \texttt{GoFish} Python package \citep{GoFish2019JOSS}, with disk-specific parameters including stellar mass, inclination, position angle, systemic velocity, and emission height adjusted to encompass all detected line emission \footnote{The stellar mass adopted for Keplerian mask construction differs
from the value listed in Table~\ref{tab:source_prop} for five targets: for
four exoALMA sample (AA~Tau, DM~Tau, LkCa~15, and SY~Cha) we adopt the
dynamical mass from \citet{Izquierdo2025}; for CI~Tau we use
$M_\star = 1.1\,M_\odot$, more consistent with the dynamical mass reported in
\citet{Simon2019}.}.
For the remaining 14 disks exhibiting non-Keplerian features, masks were instead defined by manually selected velocity ranges chosen to capture the prominent emission on a source-by-source basis. For visualization purposes, velocity maps were created with varying sigma-clipping parameters for individual targets.

Among the 37 disks we classify as dominated by Keplerian rotation, several show asymmetric emission between the blue- and red-shifted sides likely due to foreground cloud absorption. In these cases, one side is typically less affected, allowing us to still measure a disk size from the unaffected side and retain the Keplerian classification. We also note that RNO 90 shows extended emission on the blue-shifted side, to the northwest, whose origin and extent are difficult to confirm with our current data.

In contrast to the Keplerian-dominated disks, 14 disks display significantly more complex velocity structures (8 examples shown in Figure~\ref{fig:cont-co-panels-example}). We briefly characterize these systems according to their most prominent morphological features,
and a firmer classification would require a more detailed kinematic analysis, which is beyond the scope of this work. Three binary systems, VW Cha, RW Aur, and AS 205, all with projected separations within 1$\farcs$0, exhibit disrupted gas envelopes and/or large-scale tidal tails, characteristic of strong binary interactions (see detailed analyses for RW Aur in \citealt{Rodriguez2018} and AS 205 in \citealt{Kurtovic2018}).
Three low-inclination disks, DR Tau, FZ Tau, and RU Lup, display large-scale spiral features in the outer region; the spirals in DR Tau and RU Lup have been studied in detail by \citet{Huang2023_DRTau} and \citet{Huang2020_RULup}, respectively.
VZ Cha and WSB 52 show shell-like features, while Elias 2-20, HP Tau, and Haro 6-33 exhibit V-shaped structures centered at the stellar position and perpendicular to the disk orientation; both are suggestive of outflow activity, consistent with similar outflow morphologies seen in the MIR H$_2$ distribution of these five disks \citep{Narang2026}.
Some of these assignments are tentative. The CO emission of HP Tau is chaotic and coincides in scale with the ambient material seen in scattered light \citep{Garufi2024} and the associated reflection nebula \citep{Gupta2023}, which may also indicate cloud interaction. The large-scale ($\sim$8$''$) shell of WSB 52 has been attributed to a jet-driven bubble interacting with the disk \citep{Aizawa2025}.
Elias 2-27 displays extended asymmetric emission suggesting ongoing interaction with the surrounding molecular cloud, while Elias 2-24 and DG Tau show heavily truncated CO gas disks with strong nearby cloud emission.

\subsection{Dust and Gas Disk Sizes} \label{sec:disk size}
We estimated disk sizes following the approach in \citet{Long2022gas}, which used the widely adopted concept of effective radius, the radius enclosing a fraction of the total flux \citep[e.g.,][]{Tripathi2017, Andrews2018_Lmm, Sanchis2021}. The total flux is taken as the plateau value of the cumulative radial intensity profile. In this paper, we define the disk size for a given tracer as the 90\% effective radius ($R_{90\%}$) based on the azimuthally averaged, deprojected radial intensity profile. This choice ensures that most of the disk emission is included. Statistical uncertainty in disk size is estimated using bootstrap resampling with 1000 iterations. For each iteration, we generated a perturbed intensity profile by adding Gaussian noise to the intensity at each radial bin, where the noise level is taken as the 1$\sigma$ scatter divided by the square root of the number of beams along the annulus.

For the millimeter continuum emission, we measured both $R_{mm}^{img}$ from the deprojected image profile and $R_{mm}^{frank}$ from \texttt{frank} model profile (see Figure~\ref{fig:cont_rp}). Figure~\ref{fig:appendix-size} in the Appendix compares $R_{mm}^{img}$ and $R_{mm}^{frank}$ of our sample, which spans a wide range of $\sim10-360\,$au. The two measurements are generally in good agreement, with a median fractional difference of $(|R_{mm}^{frank} - R_{mm}^{img}|)/R_{mm}^{img} \approx 2.5\%$ for disks larger than 40~au. For compact disks ($R_{mm}^{img} < 40$~au), however, the fractional difference can reach 20--40\%, with the largest discrepancies found in sources that are only marginally resolved, with $R_{mm}^{img}/\theta_{\rm mm} \lesssim 2$, where $\theta_{\rm mm}$ is the average beam FWHM of continuum image (in unit of AU). In these cases, the image-plane size is likely overestimated due to beam convolution, and we therefore adopt $R_{mm}^{frank}$ as our fiducial continuum disk size where available.

For the CO gas emission, size measurements are limited to the 37 disks exhibiting well-defined Keplerian rotation patterns, with radial profiles extracted from the deprojected moment 0 maps. A subset of disks shows blue- and red-shifted asymmetries, primarily attributed to light-to-moderate foreground absorption. In these cases, the radial profiles are derived from the side of the disk that remains largely unaffected, which may lead to underestimation of disk size (see also \citealt{Long2022gas} for a discussion of the impact of cloud absorption on size measurement).
We then applied Gaussian beam deconvolution to obtain beam-corrected gas disk sizes, $R_{CO}^{deconv} = \sqrt{(R_{CO}^{img})^2 - \theta_{\rm CO}^2}$, though the correction is modest for most sources as the majority of CO-detected sources are well resolved with $R_{CO}^{img}/\theta_{\rm CO} \gtrsim 3$.

\begin{figure*}[!t]
  \includegraphics[width=0.95\linewidth]{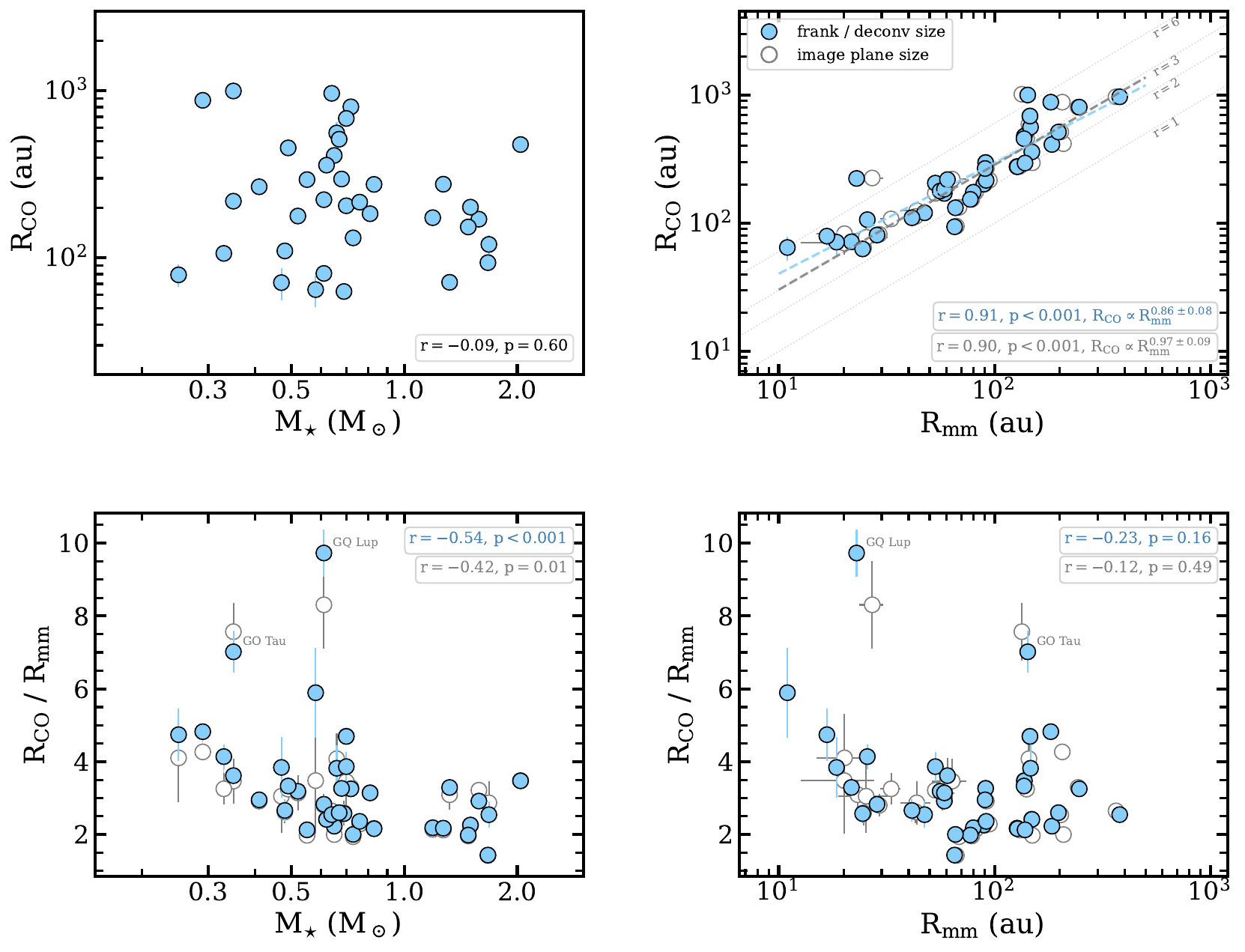}
  \caption{Gas-to-dust disk size comparison for the 37 disks where CO gas sizes can be reliably measured. \textit{Upper left:} Beam-deconvolved CO gas disk size versus stellar mass. \textit{Upper right:} CO gas versus millimeter continuum disk size, with filled blue and open circles representing quantities derived from beam correction and image plane, respectively. The dotted diagonal lines indicate constant size ratios of 1, 2, 3, and 6; the dashed lines show the \texttt{linmix} best-fit correlations of image-plane (grey) and beam-deconvolved (blue) sizes. \textit{Lower panels:} Gas-to-dust size ratio as a function of stellar mass (left) and continuum disk size (right). The Spearman coefficient $r$ and the corresponding $p$ value are annotated in each panel.}
  \label{fig:main-size}
\end{figure*}

The beam-deconvolved CO gas disk sizes span a wide range of $\sim$60--1000~au with a median value of $\sim$216~au. We find no significant correlation between CO gas disk size and stellar mass (Figure~\ref{fig:main-size}, upper left), with a Spearman rank correlation coefficient of $r = -0.09$ ($p = 0.60$). This absence of a correlation is likely driven by our sample selection, which focuses primarily on T Tauri stars. Restricting to the narrow stellar mass range of $\sim$0.5--0.8\,$M_\odot$, sources can span nearly the full range of CO disk sizes, highlighting the large intrinsic diversity of gas disk properties. We also note that the most extended disks ($R_{\rm CO} \gtrsim 500$~au) are preferentially found around the lower-mass stars in the sample.
This likely reflects a selection effect at the high-mass end of our sample, as several of the mm-brightest disks known around Herbig~Ae stars were not part of the parent JWST/MIRI Cycle~1 target list (see the right panel of Figure~\ref{fig:sample_hrd}).

In contrast, the CO gas and millimeter continuum sizes are strongly correlated (Figure~\ref{fig:main-size}, upper right), with $r = 0.91$ ($p < 0.001$), consistent with previous studies \citep{Ansdell2018, Sanchis2021, Long2022gas}. Using the Bayesian linear regression method of \citet{Kelly2007}, implemented via \texttt{linmix} in \texttt{Python}, we derive a best-fit relation of $\log R_{\rm CO} = (0.86 \pm 0.08)\log R_{\rm mm} + (0.74 \pm 0.15)$, with an intrinsic scatter of only 0.16~dex. When using image-plane measurements instead, the power-law slope becomes closer to unity ($0.97 \pm 0.09$), in better agreement with \citet{Long2022gas}, who also adopted image-plane sizes. This difference is likely due to beam convolution, which inflates the continuum sizes of marginally resolved compact sources. We note that the sample in \citet{Long2022gas} is also different from our sample here, as they probe a broader stellar mass range, including a number of very low-mass stars ($M_* \lesssim 0.2\,M_\odot$).

The gas-to-dust size ratio $R_{\rm CO}/R_{\rm mm}$ ranges from 1.4 to 9.7 across our sample, with a median of 2.95, consistent with the average value of $2.9 \pm 1.2$ reported by \citet{Long2022gas}. Seven sources have $R_{\rm CO}/R_{\rm mm} > 4$, including the two most extreme cases GO~Tau and GQ~Lup with ratios of 7.0 and 9.7, respectively.
Despite the sub-linear slope of the $R_{\rm CO}$-$R_{\rm mm}$ relation, the size ratio shows no significant correlation with continuum size ($r = -0.23$, $p = 0.16$; Figure~\ref{fig:main-size}, lower right). This null result persists after excluding the two extreme outliers.
A moderate anti-correlation with stellar mass is present ($r = -0.54$, $p < 0.001$; lower left), suggesting larger size ratios and faster radial drift around lower-mass stars \citep{Pinilla2013}. However, this trend was not found by \citet{Long2022gas} and \citet{Sanchis2021} in broader samples extending to the very low-mass star regime (M3-M6 stars), thus its significance remains uncertain (see also \citealt{Pinilla2025}).

\begin{figure*}[!t]
  \includegraphics[width=0.95\linewidth]{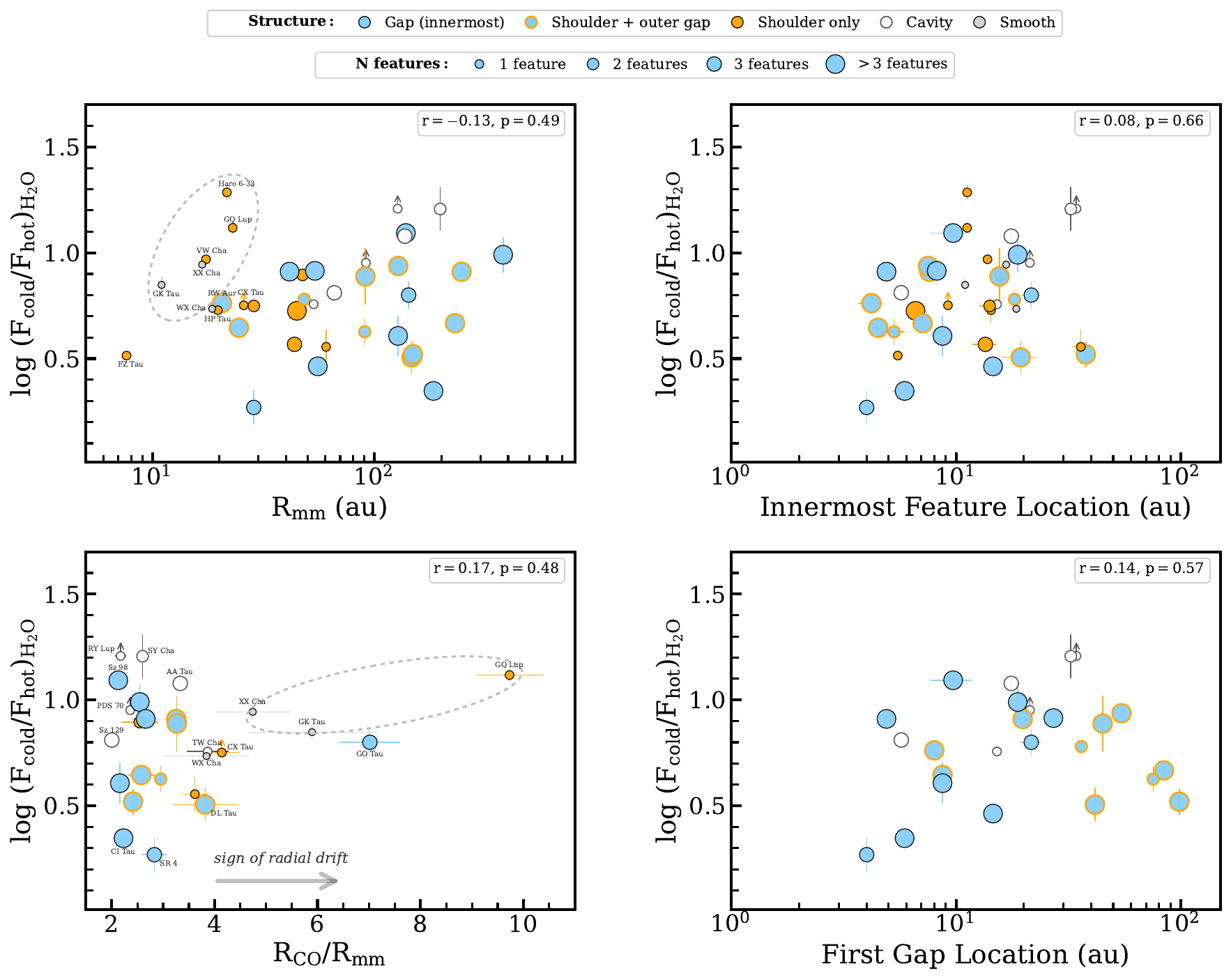}
    \caption{Cold-to-hot water flux ratio, $\log(F_{\rm cold}/F_{\rm hot})$, vs. four outer disk properties for our T~Tauri sample: mm continuum disk radius $R_{\rm mm}$ (\textit{top left}; dashed ellipse marks the group of compact smooth disks), radial location of the innermost substructure (\textit{top right}), first detected gap location (\textit{bottom right}), and gas-to-dust size ratio $R_{\rm CO}/R_{\rm mm}$ (\textit{bottom left}; dashed ellipse marks compact disks with efficient radial drift). Symbol color denotes the innermost substructure type (light blue: gaps; orange: shoulder-like features; gray filled: smooth, featureless disks; open circles: cavity disks), and symbol size scales with the number of detected substructures. Upward arrows mark lower limits on $F_{\rm cold}/F_{\rm hot}$. Spearman $r$ and $p$-value in each panel are computed for detections only, excluding cavity disks.}
  \label{fig:water-size}
\end{figure*}

\begin{figure*}[!t]
  \includegraphics[width=0.95\linewidth]{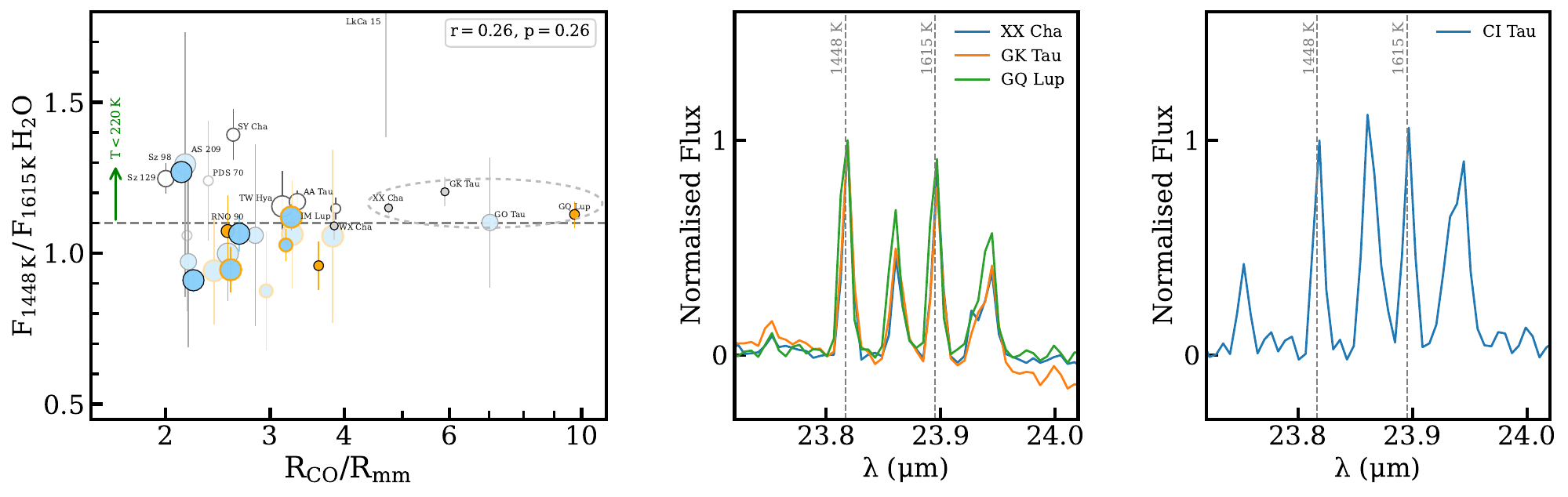}
  \caption{\textit{Left:} The 23\,$\mu$m water line ratio $F_{1448\,\rm K}/F_{1615\,\rm K}$ (a tracer of the cold component temperature) vs.\ $R_{\rm CO}/R_{\rm mm}$, restricted to sources with both lines detected at ${\rm SNR}>2.5$. Symbols follow Figure~\ref{fig:water-size}; sources with $F_{1448\,\rm K}/F_{1615\,\rm K}$ uncertainty $>0.15$ are shown in a paler shade of their marker color. The horizontal dashed line at 1.1 marks the approximate threshold above which near-snowline-temperature water ($T<220$\,K) contributes enough to be detected. \textit{Middle and Right:} Examples of continuum-subtracted MIRI/MRS spectra centering around the cold water component, normalized to the peak of the $E_{\rm up}=1448$\,K line.}
  \label{fig:water-asymmetry}
\end{figure*}

\section{Linking Outer and Inner Disk Properties} \label{sec:link}
The water lines covered by JWST/MIRI span a wide range of $E_{\rm up}$ and thus probe a broad range of excitation conditions in the disk surface layers. Higher-$E_{\rm up}$ lines often trace hotter gas located closer to the star, while lower-$E_{\rm up}$ transitions originate from cooler material at larger radii. Multiple independent lines of evidence emerge to support the picture that water emission is radially stratified, with line excitation tracing a temperature gradient across the inner disk: multi-temperature slab models are generally required to reproduce the full rotational spectrum of H$_2$O in the MIR wavelength range \citep[e.g.,][]{Temmink2024,Romero-Mirza2024}, consistent with 2D thermochemical models \citep{Vlasblom2025}; ground-based high-resolution spectroscopy revealed a gradient in line widths with $E_{\rm up}$ \citep{Banzatti2023a}, which has been recently extended to long wavelengths with MIRI/MRS \citep{Banzatti2025}.

In this work, we use two complementary diagnostics to probe the connection between the inner disk water reservoir and outer disk properties. The first is the cold water enrichment diagnostic as defined in \citet{Banzatti2025}, the flux ratio $F_{\rm cold}/F_{\rm hot}$ between water lines at $E_{\rm up}$ of $\sim$1500 and 6000\,K, tracing cold and hot gas in the inner disk, respectively. As in the original definition, we take the isolated line at 17.324\,$\mu$m to represent the hot water component. The cold flux is taken as the sum of the two low-energy lines at similar $E_{\rm up}$ (1448 and 1615\,K) near 23.85\,$\mu$m, which trace cooler water vapor closer to the water snow line.
A higher $F_{\rm cold}/F_{\rm hot}$ ratio indicates a larger cold water vapor reservoir relative to the hot component, which is often interpreted as a signature of enhanced inward delivery of icy pebbles from the outer disk that sublimate near the snowline \citep{Banzatti2023,Krijt2025}. The second diagnostic is the flux ratio of the two individual $\sim$1500\,K lines near 23.85\,$\mu$m ($F_{1448}/F_{1615}$, hereafter the 23\,$\mu$m asymmetry). LTE slab models show that $F_{1448}/F_{1615} \approx 1$ for gas at $\sim$320\,K for typical disk column densities, and the ratio rises above unity (especially in the optically thin limit) as the temperature drops to $\sim$170--200\,K, characteristic of the water snowline \citep{Banzatti2025}. While $F_{\rm cold}/F_{\rm hot}$ measures the relative amount of cold water, $F_{1448}/F_{1615}$ reveals the temperature of the cold component.

Reliable measurements of $F_{\rm cold}/F_{\rm hot}$ are derived for 39 disks in our sample\footnote{SR~21, HD~143006, and DG~Tau are excluded because both the cold and hot line fluxes are unconstrained. MY~Lup, LkCa~15, DN~Tau, TW~Hya, and DM~Tau are excluded because their flux ratio uncertainties exceed the measured value. HD~142666, HD~163296, and HD~135344B are affected by strong residual fringes in their MIRI spectra. HT~Lup is excluded because its close binary companion is unresolved by MIRI, making it impossible to reliably attribute the observed flux to individual disks.}, including 3 sources (CX Tau and two large cavity disks of PDS~70 and RY~Lup) for which only lower limits could be placed due to non-detection of the hot water component.
Among the 36 detections, $F_{\rm cold}/F_{\rm hot}$ spans from 1.9 to 19.3 with a median of $\sim$6.0. Transition disks (6 out of 39) have slightly higher median value of $F_{\rm cold}/F_{\rm hot}$ (a Kaplan-Meier estimated median of $\sim$12.0, accounting for lower limits), but not too distinct from the full disk population.
We note that 6 out of the 10 systems with $M_{\star}>1\,M_{\odot}$ in our sample are completely excluded from the water analysis (see footnote), so the discussion below concerns predominantly T Tauri stars.

Figure~\ref{fig:water-size} compares $F_{\rm cold}/F_{\rm hot}$ to outer disk properties, including millimeter continuum disk size, the innermost substructure properties, and CO-to-mm size ratio. Because the innermost feature in many disks manifests only as a shallow shoulder-like inflection point rather than a resolved gap, we also include a dedicated panel for the first gap location.
No statistically significant global correlation is found in any of the four panels, whether or not cavity disks are included. Disks spanning the full range of $F_{\rm cold}/F_{\rm hot}$ are found at similar $R_{\rm mm}$ and $R_{\rm CO}/R_{\rm mm}$, indicating that the cold water reservoir is not simply regulated by a single outer disk property.

Despite the absence of global correlations, the subset of targets with $R_{\rm CO}/R_{\rm mm} > 4$ stands out, with a higher median $F_{\rm cold}/F_{\rm hot}$ than sources below the threshold (7.9 vs. 4.4; Mann-Whitney $p = 0.12$, $N = 4$ vs.\ 15, excluding lower limits). This threshold is often considered indicative of efficient radial pebble drift \citep{Trapman2019, Toci2021}.
The difference is not statistically significant though, likely owing to the small number of sources with well-constrained $R_{\rm CO}/R_{\rm mm} > 4$; nonetheless, this group shows a clear absence of low $F_{\rm cold}/F_{\rm hot}$ values. Among the five disks with $R_{\rm CO}/R_{\rm mm} > 4$, CX~Tau hosts a water-poor spectrum, with its flux ratio only constrained as a lower limit, and GO~Tau is unusually extended with multiple gaps and rings. The remaining three sources (GQ~Lup, XX~Cha, and GK~Tau) have small dust disk radii (R$_{\rm mm} < 30$\,au) and lack strong disk substructures. Two additional compact disks (Haro~6-33 and VW~Cha) occupy a similar region in the upper-left panel of Figure~\ref{fig:water-size} (see the dashed ellipse) but lack CO size measurements, due to complex gas environment.

We next use the 23\,$\mu$m asymmetry to test if the cold water component is also cold enough to originate near the surface snowline. Figure~\ref{fig:water-asymmetry} shows $F_{1448}/F_{1615}$ as a function of $R_{\rm CO}/R_{\rm mm}$ for sources with both lines detected at ${\rm SNR} \geq 2.5$ (29 disks with both measurements). We adopt $F_{1448}/F_{1615} > 1.1$ as the threshold, above which near-snowline-temperature water ($T < 220$\,K, assuming a column density of $10^{17}$\,cm$^{-2}$, with lower temperature for higher column density) is detected \citep{Banzatti2025}. This threshold is an empirical choice, adopted to separate the drift-dominated sources from the bulk of the full disks.
Excluding cavity disks, we find no significant correlation with $R_{\rm CO}/R_{\rm mm}$ (Spearman $r=0.26$, $p=0.26$). The three compact, drift-dominated disks GQ~Lup, XX~Cha, and GK~Tau all lie above this threshold, and Haro 6-33, the other compact disk with the largest $F_{\rm cold}/F_{\rm hot}$ also shows the largest 23\,$\mu$m asymmetry in the sample.
Below $R_{\rm CO}/R_{\rm mm}=4$, most full disks fall below the threshold; the clearest exception is Sz~98, an extended, structured disk with a large asymmetry, where a recent accretion outburst may have enhanced the cold reservoir \citep{Smith2025}.

\begin{deluxetable*}{clccccc}[!t]
\tabletypesize{\scriptsize}
\tablecaption{Inner Disk Properties from JWST \label{tab:inner_disk}}
\tablewidth{0pt}
\tablehead{
\colhead{ID} & \colhead{Name} & \colhead{$n_{13-26}$} &
\colhead{$\log L_{\rm cold}$} & \colhead{$\log L_{\rm hot}$} &
\colhead{$\log(F_{\rm cold}/F_{\rm hot})$} & \colhead{23\,\micron\ asymmetry} \\
\colhead{} & \colhead{} & \colhead{} &
\colhead{($L_\odot$)} & \colhead{($L_\odot$)} &
\colhead{} & \colhead{}
}
\startdata
1 & AA Tau & $-0.53$ & $-4.60\pm0.01$ & $-5.67\pm0.03$ & $1.08\pm0.03$ & $1.17\pm0.04$ \\
2 & AS 205 N & $0.03$ & $-4.08\pm0.03$ & $-4.81\pm0.03$ & $0.73\pm0.04$ & $1.19\pm0.12$ \\
3 & AS 209 & $-0.04$ & $-5.16\pm0.09$ & $-5.77\pm0.03$ & $0.61\pm0.10$ & $1.29\pm0.44$ \\
4 & BP Tau & $-0.11$ & $-4.91\pm0.02$ & $-5.82\pm0.03$ & $0.91\pm0.03$ & $1.06\pm0.06$ \\
5 & CI Tau & $-0.18$ & $-5.03\pm0.01$ & $-5.37\pm0.02$ & $0.35\pm0.02$ & $0.91\pm0.04$ \\
6 & CX Tau & $0.00$ & $-6.10\pm0.08$ & $<-6.85$ & $>0.75$ & $<0.66$ \\
7 & CY Tau & $-1.16$ & $-5.91\pm0.02$ & $-6.47\pm0.08$ & $0.56\pm0.08$ & $0.96\pm0.08$ \\
8 & DG Tau & $-0.40$ & $<-4.90$ & $<-5.56$ & \nodata & \nodata \\
9 & DL Tau & $-0.57$ & $-5.22\pm0.06$ & $-5.72\pm0.05$ & $0.51\pm0.08$ & $1.06\pm0.29$ \\
10 & DM Tau & $1.75$ & $-6.09\pm0.12$ & $<-7.44$ & $>1.35$ & $0.50\pm0.23$ \\
11 & DN Tau & $-0.21$ & $-5.78\pm0.01$ & $<-7.12$ & $>1.33$ & $1.03\pm0.05$ \\
12 & DoAr 25 & $0.43$ & $-5.58\pm0.05$ & $-6.10\pm0.03$ & $0.52\pm0.06$ & $0.94\pm0.18$ \\
13 & DoAr 33 & $-0.87$ & $-5.67\pm0.02$ & $-6.31\pm0.03$ & $0.65\pm0.03$ & $0.95\pm0.07$ \\
14 & DR Tau & $-0.19$ & $-4.02\pm0.01$ & $-4.80\pm0.02$ & $0.78\pm0.03$ & $1.04\pm0.04$ \\
15 & Elias 2-20 & $-0.70$ & $-4.49\pm0.01$ & $-5.40\pm0.01$ & $0.91\pm0.01$ & $1.06\pm0.03$ \\
16 & Elias 2-24 & $-0.62$ & $-4.10\pm0.01$ & $-5.03\pm0.01$ & $0.94\pm0.02$ & $1.08\pm0.02$ \\
17 & Elias 2-27 & $-0.42$ & $-4.93\pm0.01$ & $-5.60\pm0.01$ & $0.67\pm0.02$ & $1.07\pm0.05$ \\
18 & FZ Tau & $-0.85$ & $-4.58\pm0.01$ & $-5.09\pm0.02$ & $0.51\pm0.02$ & $0.95\pm0.03$ \\
19 & GK Tau & $-0.08$ & $-4.87\pm0.01$ & $-5.71\pm0.04$ & $0.85\pm0.04$ & $1.20\pm0.05$ \\
20 & GO Tau & $0.19$ & $-6.18\pm0.06$ & $-6.98\pm0.03$ & $0.80\pm0.06$ & $1.10\pm0.21$ \\
21 & GQ Lup & $-0.18$ & $-4.58\pm0.01$ & $-5.69\pm0.02$ & $1.12\pm0.03$ & $1.13\pm0.04$ \\
22 & GW Lup & $-0.24$ & $-5.77\pm0.05$ & $-6.39\pm0.03$ & $0.63\pm0.06$ & $0.88\pm0.20$ \\
23 & Haro 6-33 & $0.73$ & $-4.68\pm0.02$ & $-5.96\pm0.03$ & $1.29\pm0.04$ & $1.33\pm0.09$ \\
26 & HD 143006 & $1.55$ & $<-6.00$ & $<-6.75$ & \nodata & \nodata \\
28 & HP Tau & $0.36$ & $-4.94\pm0.05$ & $-5.66\pm0.04$ & $0.73\pm0.06$ & $1.16\pm0.19$ \\
30 & IM Lup & $-0.07$ & $-4.83\pm0.01$ & $-5.74\pm0.02$ & $0.91\pm0.02$ & $1.12\pm0.04$ \\
31 & LkCa 15 & $0.73$ & $-5.79\pm0.07$ & $<-7.68$ & $>1.88$ & $1.94\pm0.56$ \\
32 & MY Lup & $0.40$ & $-5.90\pm0.08$ & $<-7.65$ & $>1.75$ & $0.97\pm0.28$ \\
33 & PDS 70 & $-0.21$ & $-6.42\pm0.05$ & $<-7.38$ & $>0.95$ & $1.24\pm0.20$ \\
34 & RNO 90 & $-0.24$ & $-4.15\pm0.03$ & $-5.04\pm0.01$ & $0.90\pm0.03$ & $1.07\pm0.12$ \\
35 & RU Lup & $0.07$ & $-4.46\pm0.02$ & $-4.92\pm0.03$ & $0.46\pm0.03$ & $1.02\pm0.07$ \\
36 & RW Aur & $-0.32$ & $-4.27\pm0.02$ & $-5.03\pm0.01$ & $0.76\pm0.02$ & $1.04\pm0.08$ \\
37 & RY Lup & $0.65$ & $-5.27\pm0.07$ & $<-6.48$ & $>1.21$ & $1.06\pm0.25$ \\
38 & SR 21 & $2.38$ & $<-4.95$ & $<-6.10$ & \nodata & \nodata \\
39 & SR 4 & $0.77$ & $-5.29\pm0.08$ & $-5.56\pm0.02$ & $0.27\pm0.08$ & $1.06\pm0.30$ \\
40 & SY Cha & $-0.09$ & $-5.01\pm0.02$ & $-6.22\pm0.10$ & $1.21\pm0.10$ & $1.39\pm0.08$ \\
41 & Sz 129 & $0.67$ & $-5.12\pm0.01$ & $-5.93\pm0.01$ & $0.81\pm0.02$ & $1.25\pm0.05$ \\
42 & Sz 98 & $-0.58$ & $-4.60\pm0.00$ & $-5.69\pm0.03$ & $1.09\pm0.03$ & $1.27\pm0.02$ \\
43 & TW Cha & $0.21$ & $-4.69\pm0.01$ & $-5.45\pm0.02$ & $0.76\pm0.02$ & $1.15\pm0.04$ \\
44 & TW Hya & $1.20$ & $-5.44\pm0.03$ & $<-7.51$ & $>2.07$ & $1.16\pm0.12$ \\
45 & V1094 Sco & $-0.73$ & $-5.94\pm0.04$ & $-6.93\pm0.07$ & $0.99\pm0.08$ & $1.00\pm0.16$ \\
46 & VW Cha & $-0.21$ & $-3.86\pm0.01$ & $-4.83\pm0.01$ & $0.97\pm0.01$ & $1.08\pm0.03$ \\
47 & VZ Cha & $-1.01$ & $-4.68\pm0.00$ & $-5.25\pm0.02$ & $0.57\pm0.02$ & $1.09\pm0.01$ \\
48 & WaOph 6 & $-0.21$ & $-5.04\pm0.05$ & $-5.93\pm0.13$ & $0.89\pm0.13$ & $1.06\pm0.18$ \\
49 & WSB 52 & $-0.22$ & $-4.36\pm0.01$ & $-5.11\pm0.02$ & $0.75\pm0.02$ & $1.06\pm0.04$ \\
50 & WX Cha & $-0.82$ & $-4.63\pm0.01$ & $-5.36\pm0.02$ & $0.74\pm0.02$ & $1.09\pm0.05$ \\
51 & XX Cha & $-0.04$ & $-4.96\pm0.01$ & $-5.91\pm0.04$ & $0.94\pm0.04$ & $1.15\pm0.02$ \\
\enddata
\tablecomments{Measurements for HD 163296, HD 142666, HD 135344B, and HT Lup were not computed for noted reasons.}
\end{deluxetable*}

\section{Discussion} \label{sec:diss}

\subsection{Outer disk dynamics and inner disk water emission}

\begin{figure*}[!t]
  \includegraphics[width=0.95\linewidth]{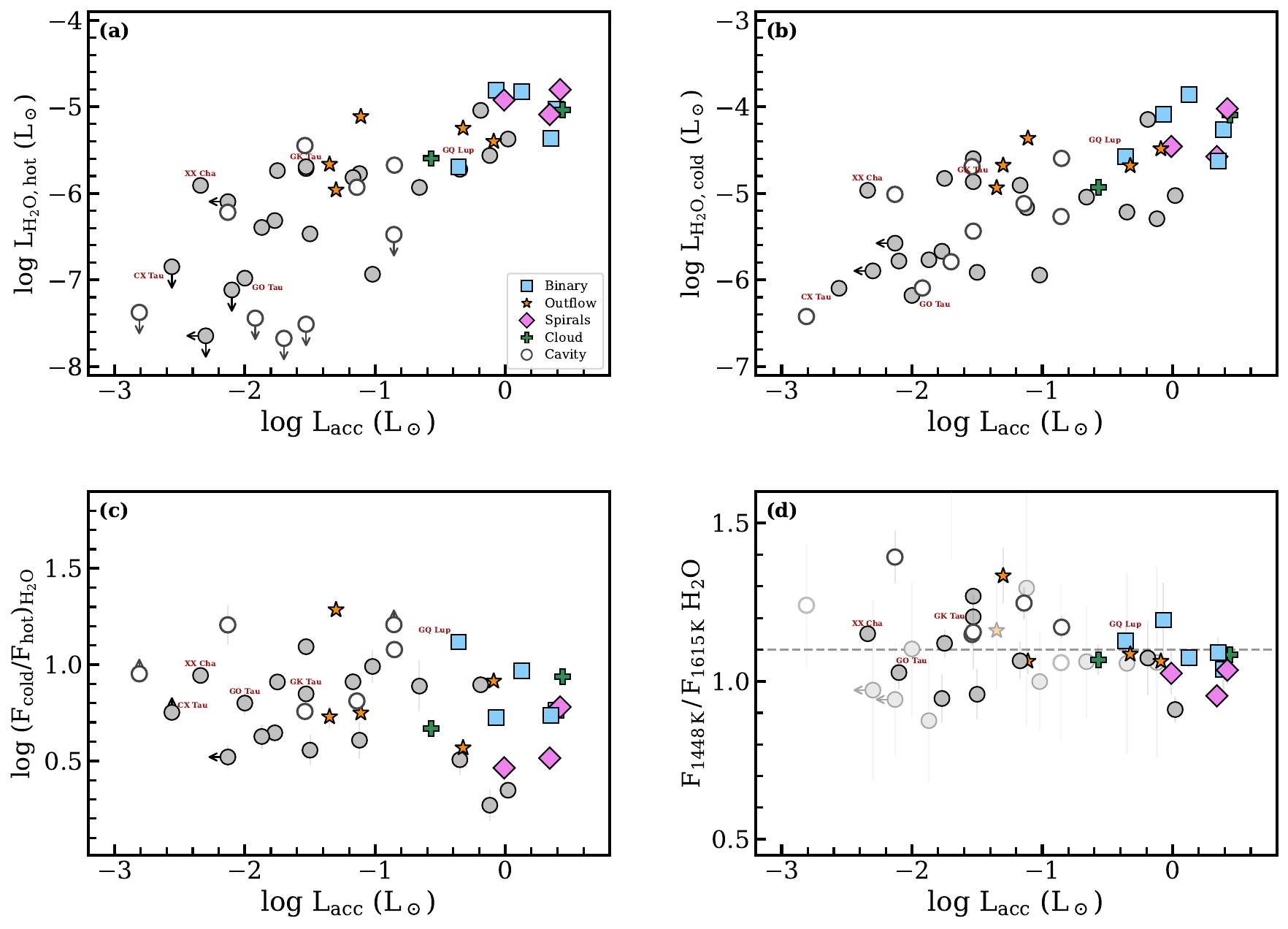}
  \caption{Hot water line luminosity (\textit{a}), cold water line luminosity (\textit{b}), cold-to-hot flux ratio (\textit{c}), and  23\,$\mu$m asymmetry ratio (\textit{d}), as a function of accretion luminosity. In all panels, marker shapes indicate the stellar/disk environment: stars for sources with outflows, squares for binary systems, diamonds for sources with spirals, plus symbols for sources with cloud contamination, open circles indicate cavity disks, and full grey dots for full normal disks. In panel (d), paler symbols mark sources with $F_{1448\,\rm K}/F_{1615\,\rm K}$ uncertainty $>0.15$.}
  \label{fig:water-lacc}
\end{figure*}

One of the strongest trends observed in mid-IR water emission from protoplanetary disks is the correlation between line luminosity and stellar accretion luminosity $L_{\rm acc}$ for both young and old disks \citep{Salyk2011, Banzatti2020, Xie2026}. We recover this correlation for both the hot and cold water components (Figure~\ref{fig:water-lacc}\textit{a, b}), reflecting the overall scaling of inner disk molecular emission with accretion-driven heating and irradiation \citep{Calahan2026}.
Line luminosities are computed from the measured line fluxes and individual source distances (see Table~\ref{tab:inner_disk}).
The hot component shows a wider dynamical range, with more non-detections concentrated at low $L_{\rm acc}$, consistent with hot water tracing the high-temperature zone set directly by accretion heating close to the star. The cold component instead remains detected down to the lowest $L_{\rm acc}$ in our sample, indicating that its emission is additionally sensitive to other physical processes \citep{Banzatti2025}. Taking the ratio $F_{\rm cold}/F_{\rm hot}$ largely removes this common $L_{\rm acc}$ dependence, especially for full disks (Figure~\ref{fig:water-lacc}\textit{c}). The 23\,$\mu$m asymmetry $F_{1448}/F_{1615}$ shows no significant dependence on $L_{\rm acc}$ (Figure~\ref{fig:water-lacc}\textit{d}). Neither diagnostic shows a strong correlation even after normalizing $L_{\rm acc}$ by stellar luminosity.

With accretion luminosity effects largely accounted for in the ratio $F_{\rm cold}/F_{\rm hot}$, the sample shows only a limited dynamical range in this ratio ($\sim$10). With this caveat in mind, in the following subsections, we examine four regimes of outer disk influence on the inner water reservoir: substructured disks with the cold water signal modulated by gap properties (Section~\ref{sec:diss_gap}); compact drift-dominated disks with efficient pebble drift (Section~\ref{sec:diss_drift}); disks with large mm cavities (Section~\ref{sec:diss_cavity}); and sources in complex outer environments (Section~\ref{sec:diss_env}).

\subsubsection{Disk substructures and the pebble supply to the snowline}
\label{sec:diss_gap}

A connection between outer disk substructures and inner disk water emission was first suggested by \citet{Banzatti2020}, who found a modest anti-correlation between disk size and inner water line flux in Spitzer spectra. This trend was interpreted as evidence that compact, relatively featureless disks allow icy pebbles to drift inward efficiently, while large disks commonly host substructures that act as barriers to pebble transport and suppress water delivery. More recently, \citet{Krijt2025} refined this picture with JWST/MIRI data, demonstrating that the radial location of the innermost gap, rather than overall disk size, may be the primary parameter regulating pebble transport to the inner disk, as previously proposed by models \citep{Kalyaan2023}. In their sample, $F_{\rm cold}/F_{\rm hot}$ increases with the radial location of the innermost gap from $\sim$10 to 30\,au, beyond which the relation flattens with considerable scatter.

Our sample for water analysis includes nearly all disks analyzed in \citet{Krijt2025}, with the exceptions of IQ~Tau and Sz~114, and adds 10 new targets (besides cavity disks). In addition to the sample size, another key difference between our analysis and \citet{Krijt2025} lies in the characterization of disk substructures, where we derive innermost gap locations consistently from visibility-plane fitting, which enables the detection of features at smaller radii than accessible in image-plane analyses. As a result, the closest substructures in several disks shift inward from the previously reported $\sim$10--20\,au down to 4--5\,au. For BP Tau, a disk with high $F_{\rm cold}/F_{\rm hot}$, we identify a deep gap at $\sim$3-5\,au with high-resolution data, whereas \citet{Krijt2025} adopted the lower resolution continuum data from \citet{Long2019} with an innermost feature only at 22\,au.
With this revised and expanded substructure census, our sample shows no global correlation between $F_{\rm cold}/F_{\rm hot}$ and innermost gap location, in contrast to the trend reported by \citet{Krijt2025}.

The rising trend found by \citet{Krijt2025} can be qualitatively understood within the framework of dust evolution models \citep[e.g.,][]{Kalyaan2023}.
In the absence of gaps, icy-rich pebbles drift unimpeded across the snowline, producing a strong cold water vapor enrichment episode that then fades as the pebble reservoir is depleted and vapor accretes onto the star. When a deep gap outside the water snowline is present, the inner water vapor enrichment is set primarily by the icy pebble reservoir between the gap and the snowline.
For a gap located farther from the star, both the enclosed pebble mass and the drift timescale increase, sustaining elevated vapor enrichment for a longer period. The fiducial models of \citet{Krijt2025} successfully reproduce this rising trend under the assumption of a strong, idealized trap that effectively blocks pebbles. Their broader model grid also explores a range of gap properties and already shows that very leaky gaps, or gaps spanning a wide range of leakiness, erode this rising trend, consistent with the larger scatter found at $>$\,30\,au in \citet{Krijt2025} and the null correlation we find above.

Gap location alone is therefore not expected to be a sufficient predictor of the inner disk water content. Recent full dust-coagulation models by \citet{Luo2026} and \citet{Houge2026} demonstrate that gap leakiness is not a fixed property of a substructure but depends on gap location as well as dust evolution parameters such as fragmentation velocity and turbulence strength. Thus, a much wider range of outcomes of water enrichment at fixed gap radius can be produced than what idealized-trap models predict.
Grain size adds a further complication as smaller grains drift more slowly and deposit volatiles less readily \citep{Birnstiel2012}, while they are also better able to cross the pressure barrier and can enhance water delivery at late disk ages once larger grains have been filtered out or depleted \citep{Houge2026}. Water vapor enrichment is also time-dependent, as the timing of gap formation and the age of the disk both affect the observable ratio at any given epoch \citep{Krijt2025}. In addition, systems that underwent recent accretion bursts can temporarily enhance the cold water reservoir and elevate $F_{\rm cold}/F_{\rm hot}$ independently of the steady-state pebble supply \citep{Houge2023, Smith2025}; this scenario may contribute to the enhanced state for Sz~98 and XX~Cha \citep{Claes2022, Smith2025, Temmink2025}.
Together, these factors could explain the large scatter we observe in $F_{\rm cold}/F_{\rm hot}$ at fixed gap radius (see also \citealt{Gasman2025}).

Despite the lack of a global correlation among our sample, a tentative rising trend in $F_{\rm cold}/F_{\rm hot}$ with innermost substructure location emerges when restricting to sources with features within ${\sim}12$\,au ($r = 0.65$, $p < 0.01$, $N = 16$; upper-right panel of Figure~\ref{fig:water-size}). This trend relies on both extremes. SR~4 and CI~Tau host deep gaps at 4--5\,au \citep[see also][]{Jennings2022a, Jennings2022b} and display the lowest $F_{\rm cold}/F_{\rm hot}$ values in the sample, consistent with near-complete pebble filtration by strong, close-in traps. On the other hand, Haro~6-33 and GQ~Lup are compact, lack strong substructures, and show the highest $F_{\rm cold}/F_{\rm hot}$ values.
When restricting to sources with a clearly resolved gap (excluding shoulder-only disks), the correlation does not reach statistical significance over the same radial range ($N = 7$; lower-right panel of Figure~\ref{fig:water-size}).
The lack of a strong dependence on gap location could arise if innermost substructures within a few au are considerably more leaky than previously assumed, as suggested by recent dust-coagulation models \citep{Houge2026}.
A recent JWST study comparing young (Taurus) and intermediate-age (IC~348) disk samples finds sustained chemical diversity across the two populations \citep{Carr2026}, suggesting instead that gaps retain moderate-to-low leakiness. Several leaky traps in series could further reduce the pebble flux reaching the snowline \citep[e.g.,][]{McClure2020}. In our sample, however, the number of detected substructures does not correlate with $F_{\rm cold}/F_{\rm hot}$ ($r=-0.16$, $p=0.39$), nor with the residuals of the tentative trend for disks with innermost features within ${\sim}12$\,au ($r=0.11$, $p=0.69$). This suggests that the depth and leakiness of individual gaps may matter more than their location and number in determining the amount of pebble transport.

There are several caveats that apply to this analysis. Systems whose innermost resolved gap lies at large radii may host unidentified substructures closer in, as many of our sample do not have sufficient resolution to resolve structures within $\sim$5-10\,au. On the other hand, visibility-plane fitting itself carries uncertainties in both the location and depth of identified features. A larger sample with ALMA data at 2-3\,au resolution paired with JWST spectra is needed to robustly establish the connection between outer disk substructures and inner disk water properties. That said, our current data show no indication that higher spatial resolution would restore a positive correlation between innermost gap location and cold water enrichment. Such a correlation would require disks with low $F_{\rm cold}/F_{\rm hot}$ to preferentially host closer-in gaps that are currently unresolved, yet disks with high and low $F_{\rm cold}/F_{\rm hot}$ in our sample have similar beam size distributions.
Meanwhile, gap depth cannot be reliably constrained from mm continuum images alone without complementary gas kinematics, making it difficult to distinguish between deep, efficient traps and shallow, leaky gaps that may produce very different water enrichment histories.

Furthermore, $F_{\rm cold}/F_{\rm hot}$ might be an imperfect tracer of the actual pebble delivery, as infrared observations probe only the water vapor above the $\tau_\nu = 1$ dust continuum surface \citep[e.g.,][]{Woitke2018}. When icy pebbles sublimate at the snowline, the released refractory grains accumulate inside the snowline as small, slowly drifting fragments, raising the $\tau_\nu = 1$ surface to higher disk layers and hiding an increasing fraction of the water vapor reservoir from view \citep{Houge2025, Sellek2025}, such that ongoing pebble delivery may go largely undetected in MIRI spectra \citep{Vlasblom2026}. This scenario assumes that bare grains are more fragile than icy ones, and also depends on the dust opacity and the level of turbulence. More broadly, the observable water column density depends not only on the pebble flux but also on the dust size distribution inside the snow line, the efficiency of UV and chemical processing, and the transport of water vapor after sublimation \citep{Kaeufer2026}. High stellar accretion rates complicate this effect by expanding the hot water emitting area and boosting the hot water signal, while some cold water becomes hidden beneath the optically thickened dust surface \citep{Calahan2026}, biasing $F_{\rm cold}/F_{\rm hot}$ downward independently of the true pebble supply.
Disentangling these degeneracies will require dust evolution models that self-consistently couple gap leakiness, grain size evolution, and the $\tau_\nu=1$ dust surface across the range of stellar and disk properties spanned by our sample, extending frameworks such as \citet{Krijt2025} and \citet{Houge2026} beyond their current setups.

\subsubsection{Cold water enrichment in compact, drift-dominated disks}
\label{sec:diss_drift}
Section~\ref{sec:link} identified GK~Tau, XX~Cha, and GQ~Lup as the compact, structureless disks anchoring the elevated $F_{\rm cold}/F_{\rm hot}$ seen among sources with $R_{\rm CO}/R_{\rm mm} > 4$ (bottom-left panel of Figure~\ref{fig:water-size}). With no strong traps standing between the outer disk and the snowline, the pebble flux reaching the inner disk in these three systems should be essentially unsuppressed, offering the clearest test of whether efficient radial drift leaves a direct imprint on the inner disk water reservoir. The 23\,$\mu$m asymmetry with $F_{1448}/F_{1615} > 1.1$ further confirms that their elevated cold water reservoir resides close to the snowline.
The elevated cold water signal thus reflects a sustained signature of unimpeded drift, consistent with the scenario of \citet{Banzatti2020, Banzatti2023}, in which continuous pebble drift replenishes the inner disk water reservoir near the snowline.
We note that GK~Tau and CI~Tau, representative of compact featureless and large structured disks, are the pair that \citet{Banzatti2023} compared to establish the observational link between pebble drift and inner disk water enrichment.
Translating this cold water reservoir into a pebble budget, \citet{Romero-Mirza2024} derived icy pebble mass fluxes into the inner disk that are more than an order of magnitude higher in GQ~Lup and GK~Tau than in the extended, structured disk CI~Tau.

GQ Lup has the largest $R_{\rm CO}/R_{\rm mm}$ in our sample and hosts notable system complexities.
It has a substellar companion GQ~Lup~B ($\sim$30\,$M_{\rm Jup}$, $a = 117\pm23$\,au, $e = 0.24^{+0.32}_{-0.17}$, companion-disk misalignment $84^\circ \pm 9^\circ$; \citealt{Stolker2021}) and a low-mass ($\sim$0.15\,$M_\odot$) wide companion GQ~Lup~C at a projected separation of $\sim$2400\,au \citep{Alcala2020}. At more than ten times the CO disk radius, GQ~Lup~C is expected to be too distant to be dynamically relevant.
The dust disk of GQ Lup is compact ($R_{\rm mm} \approx 23$\,au), while its CO gas disk extends to $\sim$220\,au, well beyond the orbit of GQ~Lup~B. This is in tension with standard binary interaction models \citep{Artymowicz1994, Miranda2015}, which predict that the circumprimary disk would truncate at 0.3-0.5$a$,
though these models do not extend to the extreme stellar mass ratio of this system ($q\sim 0.05$).
Despite the highly inclined companion orbit that could induce secular perturbations and disk warping, the CO kinematics show no strong non-Keplerian features beyond marginal ($\sim3\sigma$) deviations at the outer disk edge (see \citealt{MacGregor2017} and Figure~\ref{fig:cont-co-panels-normal-remaining}).
Even allowing for some overestimate of the gas disk extent from such distortions, the well-ordered inner gas disk remains far more extended than the dust disk, still placing GQ Lup in the drift-dominated regime.
The dynamical influence of GQ~Lup~B may in fact enhance radial drift by accelerating inward dust migration from disk truncation \citep{Zagaria2021}, a scenario invoked to explain enhanced water emission in a number of other binary disks \citep{Grant2024_DFTau, Kurtovic2026, Yao2026arXiv}. For the other two systems,
GK Tau has a wide stellar companion GI Tau at $\sim$1700\,au, which is likely too far to be dynamically relevant; XX Cha has no known companion.

CX~Tau and GO~Tau, the other two disks with $R_{\rm CO}/R_{\rm mm} > 4$ (Section~\ref{sec:link}), differ from the three above. \citet{Vlasblom2025a} proposed that the CO$_2$-dominated, H$_2$O-poor spectrum of CX Tau indicates a dust trap located between the H$_2$O and CO$_2$ snowlines that selectively suppresses water vapor delivery while allowing CO$_2$ ice to sublimate normally.
GO~Tau hosts multiple gaps and has weak water emission for its accretion luminosity. Its MIRI spectrum is very different from that of CX~Tau, with prominent C$_2$H$_2$ and normal CO$_2$ emission, which may point to a higher C/O ratio that suppresses water formation, although the origin of this difference remains unsettled (see also \citealt{Colmenares2024} for DoAr~33). Despite the weak water emission, GO~Tau has an $F_{\rm cold}/F_{\rm hot}$ comparable to GK~Tau and XX~Cha. This may indicate that its gaps are leaky, so that part of the pebbles drifting inward, as suggested by its high $R_{\rm CO}/R_{\rm mm}$, still reach the snowline. GK~Tau and XX~Cha have similarly low accretion rates, but both their hot and cold lines are more than an order of magnitude brighter than those of GO~Tau, indicating a more water-rich inner disk.

The sample with clear multi-indicator evidence remains small, in part because our JWST Cycle 1 sample is biased toward mm-bright disks, which more often host substructure; GK~Tau and XX~Cha are the two faintest mm continuum sources in our sample. A compact dust disk does not necessarily imply efficient radial drift, however (see Section~\ref{sec:diss_initial_size}), so the active drift inferred for these few examples may not be the norm among compact disks more broadly.

\subsubsection{Disks with large mm cavity}
\label{sec:diss_cavity}
In transition disks with a resolved mm dust cavity, large pebbles should have either already drifted across the water snowline before the cavity opened, or remain trapped at the pressure bump outside its edge. In either case, essentially no ongoing large-pebble flux reaches the snowline today, making these disks a special laboratory for testing how the absence of pebble inflow shapes the water reservoir of the inner disk. If cold water were regulated primarily by this large-pebble flux, as inferred for the compact, structureless disks of Section~\ref{sec:diss_drift}, these disks are expected to show depleted cold water emission.

Among our sample of 13 transition disks, about half have reliable $F_{\rm cold}/F_{\rm hot}$ constraints. Two of these, PDS~70 and RY~Lup, yield only lower limits from undetected hot water, while the remaining four (AA~Tau, SY~Cha, Sz~129, TW~Cha) have secure detections of both components. These six disks show slightly higher median $F_{\rm cold}/F_{\rm hot}$ than that of the full-disk sample, as also found in \citet{Mallaney2026} (see their Figure~6), though they do not stand out sharply from the rest (see Figure~\ref{fig:water-lacc}c).

The high $F_{\rm cold}/F_{\rm hot}$ values for PDS~70 and RY~Lup are driven mainly by their non-detection of hot water. Both are classified as ``molecule-poor'' in \citet{Mallaney2026}, which are globally subluminous in all molecular tracers, with the inner disk largely cleared of both gas and dust.  SY~Cha, Sz~129 and TW~Cha are instead ``molecule-rich'' (MR) as defined in \citet{Mallaney2026}, where an inner molecular gas reservoir persists, likely shielded from UV photodissociation by residual small $\mu$m grains. The water emission of AA~Tau resembles a MR disk, with a negative IR spectral index that indicates the presence of small grains (see Table~\ref{tab:inner_disk}).
The moderate-to-high $F_{\rm cold}/F_{\rm hot}$ values among these MR sources may therefore imply a genuine enhancement of the cold component. One possible explanation is sublimation from small icy grains that leak through the cavity in the absence of active pebble supply \citep{Pinilla2024, Houge2026, Sierra2025}. Direct irradiation on the cavity wall may have also enhanced ice sublimation.
This is consistent with the 23\,$\mu$m asymmetry. The four cavity disks with secure water detection show $F_{1448}/F_{1615}$ values exceeding the $>$1.1 threshold (Figure~\ref{fig:water-asymmetry}), confirming that cold water in these four disks genuinely resides close to the snowline.

We also note that SR 4 and HP Tau appear in \citet{Mallaney2026} as cavity disks based on their high IR spectral index (both MR), but show no clear sign of mm dust cavity at current resolution and are thus treated as full disks in our analysis.

\begin{figure}[!t]
  \centering
  \includegraphics[width=0.95\linewidth]{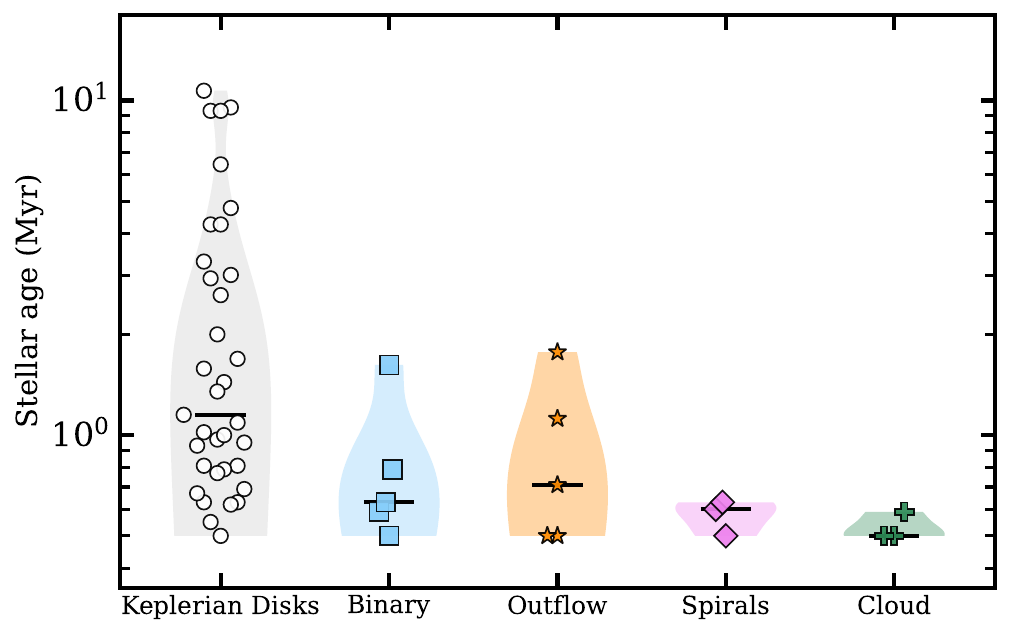}
  \caption{Comparison of stellar age distributions for disks with different outer disk gas morphologies. }
  \label{fig:age-complex}
\end{figure}

\subsubsection{Disks with complex outer environments} \label{sec:diss_env}
In Section~\ref{sec:co-disk}, we identified 14 disks that exhibit prominent non-Keplerian disk features and were excluded from the CO gas disk size measurement and the evaluation of radial drift efficiency. Figure~\ref{fig:water-lacc} highlights the inner disk water emission properties of these disks with those of the well-behaved sample.
No statistically significant differences are seen between the two groups in either water diagnostic used above ($F_{\rm cold}/F_{\rm hot}$ and 23\,$\mu$m asymmetry). Thus the complex outer environment, as defined here, does not leave a distinct imprint on the inner disk water emission.

The most prominent difference is in accretion luminosity. The complex-environment group has a median $\log L_{\rm acc} = -0.08\,L_\odot$, compared to $-$1.50 for normal disks ($p = 0.002$, excluding cavities; Figure~\ref{fig:water-lacc}). This elevated accretion is not due to stellar mass differences. Given the IR line and accretion luminosity correlations, this drives a corresponding offset in water line luminosities. Among the categories defined in Section~\ref{sec:co-disk}, disks with spirals and binary systems show the highest accretion rates and water line luminosities. The large-scale CO spiral arms in DR Tau and RU Lup have been interpreted as late-stage infall signals that deliver fresh materials to the disk and sustain anomalously high accretion into the Class~II stage \citep{Huang2020_RULup, Huang2023_DRTau}.

Figure~\ref{fig:age-complex} shows that complex-environment disks are significantly younger than the simple-environment sample (KS $p = 0.02$--$0.03$, robust to whether binaries are included), especially for those with spirals or associated with clouds. Under the framework of viscous evolution, disk accretion rates decline with stellar age \citep{Hartmann1998},
with observational support from comparisons of embedded Class~I disks to more evolved Class~II systems \citep{Fiorellino2021}. The elevated accretion across all complex-environment sub-groups is therefore consistent with, and provides independent support for, their preferential youth. On the other hand, we do not find clear evidence of cold water enrichment from rapid pebble drift that might otherwise be expected in these young systems (see recent work on young disks by \citealt{Zhang2026}).

\subsection{Initial disk size from formation}
\label{sec:diss_initial_size}

The wide range of CO gas disk sizes ($R_{\rm CO}\sim$60--1000~au) in our sample is striking given that most targets fall within a narrow range of stellar masses ($\sim$0.5--0.8~$M_\odot$) and similar ages. No significant correlation between $R_{\rm CO}$ and stellar mass is found in our sample (Section~\ref{sec:outer}), and \citet{Long2022gas} identified only a weak correlation when extending to include very low-mass stars. This diversity may reflect divergent evolutionary histories (e.g., differences in effective viscosity or wind-driven mass loss), or diversity that is largely inherited from the formation stage. Compact dust disks are commonly attributed to radial drift depleting the outer disk over time \citep[e.g.,][]{Long2019, GuerraAlvarado2025}, but the roughly constant $R_{\rm CO}/R_{\rm mm}$ across a wide range of $R_{\rm mm}$ in our sample is equally consistent with many disks having simply been born small (see also \citealt{Miotello2021}, \citealt{Ricciardi2026}); such disks would not necessarily carry the enhanced cold-water signature expected from active pebble drift.

Disk formation models offer physical grounds for this initial diversity. Prestellar cores of similar mass can carry very different angular momenta, producing a range of initial disk radii \citep{Williams2026}. Cluster-scale hydrodynamic simulations show that stochastic encounters and tidal truncation alone can also generate a wide range of disk radii among protostars of similar mass \citep{Bate2018}. In non-ideal MHD simulations, disk size is instead set by the efficiency of magnetic braking, which depends on field strength, geometry, and non-ideal effects \citep{Zhao2021, Tsukamoto2023}.
Observationally, the VANDAM Orion survey found a large spread in protostellar dust disk radii already at the Class~0/I stage \citep{Tobin2020}, consistent with size diversity being present before the Class~II phase probed here. Quantifying disk size evolution with stellar age has been proposed as a test to distinguish viscous spreading from MHD-wind-driven disk evolution (e.g., AGEPRO, \citealt{Zhang2025}), but such a trend may be difficult to detect if the initial spread at fixed stellar mass is already this broad.

\section{Summary} \label{sec:summ}
In this paper, we present a homogeneous analysis of outer disk dust and gas properties for 51 protoplanetary disks from four JWST/MIRI Cycle 1 programs, combining new ALMA observations with archival data. We characterize mm continuum substructures and CO gas emission morphologies, measure dust and gas disk sizes, and investigate how these outer disk properties regulate the inner disk water content probed by JWST/MIRI. The main results are as follows.

\begin{enumerate}

\item Visibility-plane fitting with \texttt{frank} reveals emission features in otherwise smooth-appearing disks and new features at smaller radii than accessible in image-plane analyses, leaving only GK~Tau, XX~Cha, and WX~Cha featureless at current resolution of 0.$''$05-0.$''$1.

\item The CO gas disk sizes of our sample span a wide range of $\sim$60--1000\,au (median of $\sim$216\,au). CO and mm continuum sizes are strongly correlated ($r = 0.91$), with $R_{\rm CO}/R_{\rm mm}$ ranging from 1.4--9.7 (median 2.95) and showing no dependence on stellar mass. This diversity at fixed stellar mass may be largely inherited from formation, though disk evolution cannot be excluded.

\item We find no global correlation between the cold-to-hot water flux ratio $F_{\rm cold}/F_{\rm hot}$ and any single outer disk property, including disk size, gas-to-dust disk size ratio, and innermost substructure locations. The large scatter of $F_{\rm cold}/F_{\rm hot}$ at fixed gap radius indicates that many other factors, rather than gap location alone, set the observed inner disk water content.

\item A small subset of compact ($R_{\rm mm} \lesssim 30$\,au) disks (GK~Tau, XX~Cha, and GQ~Lup), with no strong substructures and large gas-to-dust size ratios ($R_{\rm CO}/R_{\rm mm}>4$) show both high $F_{\rm cold}/F_{\rm hot}$ and near-snowline water temperatures ($F_{1448}/F_{1615} > 1.1$), consistent with ongoing, unimpeded inward delivery of icy pebbles.

\item Transition disks with resolved mm cavities show comparable or slightly higher $F_{\rm cold}/F_{\rm hot}$ to full disks, suggesting that other factors than pebble drift affect the ratio in these disks. One possible explanation is the sublimation of small icy grains leaking through the cavity in molecule-rich systems.

\item Disks in complex outer environments (outflows, spirals, binary interactions, cloud-interacting sources) show systematically higher accretion luminosities, likely reflecting their younger ages, but no distinct inner disk water emission properties.

Together, these results indicate that outer disk dust evolution shapes the inner disk water content in a complex way, with no simple dependence on any single outer disk property.
Evidence for inner water enrichment by pebble drift is clearest in the compact structureless disks, and confirming it calls for a larger, carefully selected sample of such disks. For substructured disks, progress requires constraining gap leakiness and grain properties with ALMA gas and dust observations. Moreover, tracers of pebble drift in addition to the $F_{\rm cold}/F_{\rm hot}$ line ratio should be searched for to clarify some of the existing degeneracies.

\end{enumerate}
\paragraph{Acknowledgments}
We thank the referee for careful reading of the manuscript and helpful suggestions. This work was supported by funding from the National SKA Program of China under Grant No. 2025SKA0120100. This work was also supported by the high-performance computing platform of Peking University.
Support for F.L. was provided by NASA through the NASA Hubble Fellowship grant \#HST-HF2-51512.001-A awarded by the Space Telescope Science Institute, which is operated by the Association of Universities for Research in Astronomy, Incorporated, under NASA contract NAS5-26555.
E.R. and G.R. acknowledge support from the European Union (ERC Starting Grant DiscEvol, project number 101039651). C.F.M is funded by the European Union (ERC, WANDA, 101039452). FMe received funding from the ERC under the European Union's Horizon Europe research and innovation program (grant agreement No. 101053020, project Dust2Planets). Views and opinions expressed are, however, those of the author(s) only and do not necessarily reflect those of the European Union or the European Research Council. Neither the European Union nor the granting authority can be held responsible for them. D.H. is supported by a Center for Informatics and Computation in Astronomy (CICA) grant and grant number 110J0353I9 from the Ministry of Education of Taiwan. D.H. also acknowledges support from the National Science and Technology Council, Taiwan (Grant NSTC111-2112-M-007-014-MY3, NSTC113-2639-M-A49-002-ASP, and NSTC113-2112-M-007-027). XNB is supported by National Science Foundation of China (12325304). SK and TK acknowledge support from STFC Grant ST/Y002415/1. JW is funded by the UK Science and Technology Facilities Council (STFC), grant code ST/Y509383/1.

This work includes observations made with the NASA/ESA/CSA James Webb Space Telescope. The JWST data used in this paper can be found in MAST: \dataset[10.17909/pxks-nn37]{http://dx.doi.org/10.17909/pxks-nn37}.
This paper makes use of the following ALMA data: ADS/JAO.ALMA\#2021.1.00871.S, ADS/JAO.ALMA\#2013.1.00498.S, ADS/JAO.ALMA\#2015.1.00168.S, ADS/JAO.ALMA\#2015.1.00222.S, ADS/JAO.ALMA\#2015.1.00773.S, ADS/JAO.ALMA\#2015.1.00847.S, ADS/JAO.ALMA\#2015.1.01268.S, ADS/JAO.ALMA\#2016.1.00627.S, ADS/JAO.ALMA\#2016.1.00715.S, ADS/JAO.ALMA\#2016.1.01042.S, ADS/JAO.ALMA\#2016.1.01164.S, ADS/JAO.ALMA\#2016.1.01239.S, ADS/JAO.ALMA\#2016.1.01370.S, ADS/JAO.ALMA\#2017.1.00449.S, ADS/JAO.ALMA\#2017.1.01167.S, ADS/JAO.ALMA\#2018.1.00689.S, ADS/JAO.ALMA\#2018.1.00771.S, ADS/JAO.ALMA\#2018.1.01458.S, ADS/JAO.ALMA\#2019.1.00607.S, ADS/JAO.ALMA\#2019.1.01619.S, ADS/JAO.ALMA\#2021.1.00854.S, ADS/JAO.ALMA\#2021.1.01050.S. ALMA is a partnership of ESO (representing its member states), NSF (USA) and NINS (Japan), together with NRC (Canada), MOST and ASIAA (Taiwan), and KASI (Republic of Korea), in cooperation with the Republic of Chile. The Joint ALMA Observatory is operated by ESO, AUI/NRAO and NAOJ.

\vspace{5mm}
\facilities{JWST(MIRI), ALMA}

\software{CASA \citep{CASATeam2022PASP},
          emcee \citep{ForemanMackey2013},
          ysoisochrone \citep{Deng2025JOSS},
         iSLAT \citep{Jellison2024}
          }

\clearpage
\appendix

\section{Data info} \label{sec:appendix-data}
\begin{figure*}[!t]
  \includegraphics[width=0.95\linewidth, trim=100 300 50 400, clip]{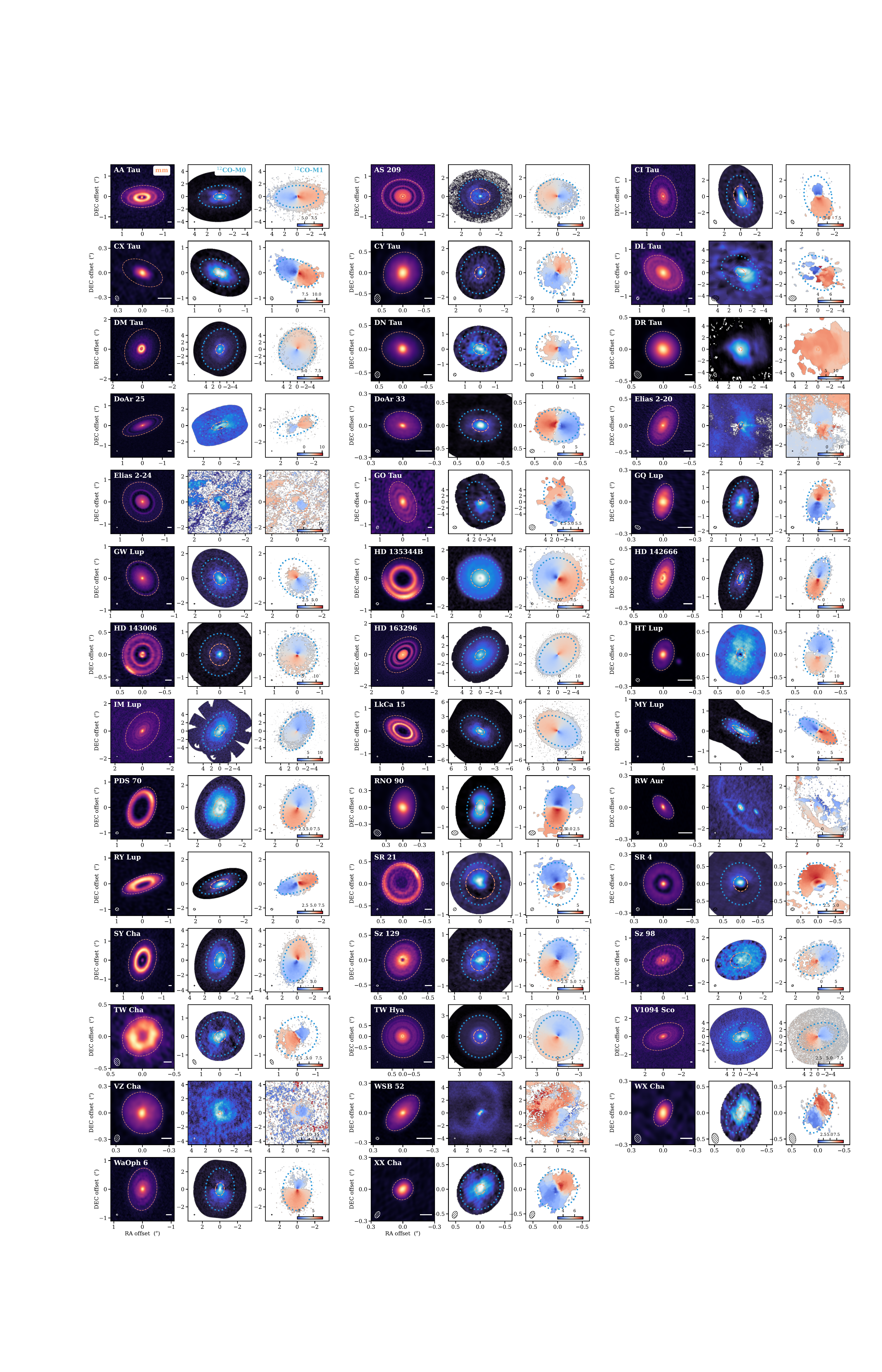}
  \caption{ALMA mm continuum and CO gas maps for the remaining 41 disks.}
  \label{fig:cont-co-panels-normal-remaining}
\end{figure*}

Details of JWST and ALMA data used in this work are summarized in Table~\ref{tab:data_info}. Figure~\ref{fig:cont-co-panels-normal-remaining} shows the mm continuum and CO maps of our sample.

\renewcommand{\arraystretch}{0.9}
\begin{deluxetable*}{ccccccccccc}
\tabletypesize{\scriptsize}
\tablecaption{Data Info\label{tab:data_info}}
\tablewidth{0pt}
\tablehead{
\colhead{Name} & \colhead{JWST PID} & \multicolumn{4}{c}{mm cont} & \multicolumn{5}{c}{CO line} \\
\cmidrule(lr){3-6} \cmidrule(lr){7-11}
\colhead{} & \colhead{} & \colhead{freq.} & \colhead{beam} & \colhead{rms} & \colhead{REF} & \colhead{transition} & \colhead{beam} & \colhead{rms} &  \colhead{$\delta$v} &  \colhead{REF}\\
\colhead{} & \colhead{} & \colhead{(GHz)} & \colhead{(mas)} & \colhead{(mJy/beam)} & \colhead{} & \colhead{} & \colhead{(mas)} & \colhead{(mJy/beam)} &  \colhead{(km/s)} &  \colhead{}\\
}
\colnumbers
\startdata
AA Tau & GO-1282 & 338 & $95 \times 75$ & 0.024 & \makecell{exoALMA} & $J=3-2$ & $150 \times 150$ & 3.44 & 0.1 & \makecell{exoALMA} \\
AS 205 N & GO-1584 & 234 & $37 \times 24$ & 0.016 & \makecell{DSHARP} & $J=2-1$ & $115 \times 92$ & 1.43 & 0.35 & \makecell{DSHARP} \\
AS 209 & GO-1584 & 239 & $38 \times 36$ & 0.017 & \makecell{DSHARP} & $J=2-1$ & $95 \times 72$ & 0.9 & 0.35 & \makecell{DSHARP} \\
BP Tau & GO-1282 & 225 & $34 \times 29$ & 0.016 & \makecell{2019.1.00607.S\\2021.1.01050.S} & $J=2-1$ & $209 \times 172$ & 2.94 & 0.4 & \makecell{2019.1.00607.S\\2021.1.01050.S} \\
CI Tau & GO-1640 & 230 & $69 \times 48$ & 0.017 & \makecell{2016.1.01370.S\\2016.1.01164.S\\2016.1.00627.S} & $J=2-1$ & $517 \times 354$ & 8.15 & 0.25 & \makecell{2016.1.00627.S} \\
CX Tau & GO-1282 & 232 & $62 \times 43$ & 0.031 & \makecell{2016.1.00715.S} & $J=2-1$ & $152 \times 119$ & 2.14 & 0.5 & \makecell{2016.1.00715.S} \\
CY Tau & GO-1282 & 225 & $196 \times 136$ & 0.108 & \makecell{2013.1.00498.S\\2021.1.00854.S} & $J=2-1$ & $266 \times 188$ & 4.73 & 0.5 & \makecell{2013.1.00498.S\\2021.1.00854.S} \\
DG Tau & GO-1282 & 230 & $39 \times 30$ & 0.012 & \makecell{2015.1.01268.S} & $J=2-1$ & $191 \times 149$ & 4.78 & 0.65 & \makecell{2016.1.01042.S} \\
DL Tau & GO-1282 & 226 & $136 \times 109$ & 0.053 & \makecell{\citet{Long2018}} & $J=2-1$ & $1319 \times 944$ & 79.59 & 0.25 & \makecell{\citet{Long2022gas}} \\
DM Tau & GO-1282 & 338 & $92 \times 77$ & 0.022 & \makecell{exoALMA} & $J=3-2$ & $150 \times 150$ & 3.11 & 0.1 & \makecell{exoALMA} \\
DN Tau & GO-1282 & 226 & $133 \times 107$ & 0.053 & \makecell{\citet{Long2018}} & $J=2-1$ & $205 \times 196$ & 5.56 & 0.65 & \makecell{2016.1.01042.S} \\
DoAr 25 & GO-1584 & 239 & $40 \times 21$ & 0.013 & \makecell{DSHARP} & $J=2-1$ & $101 \times 78$ & 1.16 & 0.35 & \makecell{DSHARP} \\
DoAr 33 & GO-1584 & 239 & $36 \times 24$ & 0.017 & \makecell{DSHARP} & $J=2-1$ & $103 \times 79$ & 1.27 & 0.35 & \makecell{DSHARP} \\
DR Tau & GO-1282 & 230 & $127 \times 99$ & 0.052 & \makecell{\citet{Long2019}} & $J=2-1$ & $786 \times 467$ & 14.06 & 0.2 & \makecell{\citet{Huang2023_DRTau}} \\
Elias 2-20 & GO-1584 & 232 & $31 \times 23$ & 0.015 & \makecell{DSHARP} & $J=2-1$ & $102 \times 72$ & 0.98 & 0.35 & \makecell{DSHARP} \\
Elias 2-24 & GO-1584 & 232 & $36 \times 34$ & 0.019 & \makecell{DSHARP} & $J=2-1$ & $94 \times 60$ & 1.21 & 0.35 & \makecell{DSHARP} \\
Elias 2-27 & GO-1584 & 232 & $48 \times 47$ & 0.015 & \makecell{DSHARP} & $J=2-1$ & $132 \times 111$ & 1.28 & 0.35 & \makecell{DSHARP} \\
FZ Tau & GO-1549 & 226 & $57 \times 43$ & 0.028 & \makecell{2021.1.00854.S} & $J=2-1$ & $244 \times 215$ & 2.24 & 0.5 & \makecell{2021.1.00854.S} \\
GK Tau & GO-1640 & 226 & $124 \times 107$ & 0.05 & \makecell{2016.1.01164.S} & $J=2-1$ & $233 \times 187$ & 3.59 & 0.3 & \makecell{2018.1.00771.S} \\
GO Tau & GO-1640 & 226 & $138 \times 105$ & 0.052 & \makecell{2016.1.01164.S} & $J=2-1$ & $1330 \times 940$ & 104.01 & 0.25 & \makecell{\citet{Long2022gas}} \\
GQ Lup & GO-1640 & 233 & $57 \times 33$ & 0.038 & \makecell{2015.1.00773.S} & $J=2-1$ & $216 \times 163$ & 5.76 & 0.5 & \makecell{2016.1.01239.S} \\
GW Lup & GO-1282 & 239 & $44 \times 42$ & 0.015 & \makecell{DSHARP} & $J=2-1$ & $109 \times 81$ & 1.35 & 0.35 & \makecell{DSHARP} \\
Haro 6-33 & GO-1640 & 226 & $61 \times 50$ & 0.037 & \makecell{2021.1.00854.S} & $J=2-1$ & $179 \times 149$ & 2.72 & 0.5 & \makecell{2021.1.00854.S} \\
HD 135344B & GO-1282 & 338 & $89 \times 76$ & 0.045 & \makecell{exoALMA} & $J=3-2$ & $150 \times 150$ & 2.59 & 0.1 & \makecell{exoALMA} \\
HD 142666 & GO-1584 & 232 & $31 \times 22$ & 0.013 & \makecell{DSHARP} & $J=2-1$ & $77 \times 61$ & 1.25 & 0.35 & \makecell{DSHARP} \\
HD 143006 & GO-1584 & 239 & $53 \times 36$ & 0.011 & \makecell{DSHARP} & $J=2-1$ & $66 \times 49$ & 1.01 & 0.32 & \makecell{DSHARP} \\
HD 163296 & GO-1584 & 239 & $47 \times 38$ & 0.023 & \makecell{DSHARP} & $J=2-1$ & $104 \times 95$ & 1.48 & 0.32 & \makecell{DSHARP} \\
HP Tau & GO-1640 & 226 & $131 \times 110$ & 0.052 & \makecell{2016.1.01164.S} & $J=2-1$ & $229 \times 200$ & 4.02 & 0.25 & \makecell{2021.1.01050.S} \\
HT Lup & GO-1584 & 239 & $38 \times 33$ & 0.014 & \makecell{DSHARP} & $J=2-1$ & $53 \times 50$ & 1.02 & 0.35 & \makecell{DSHARP} \\
IM Lup & GO-1282 & 239 & $44 \times 43$ & 0.014 & \makecell{DSHARP} & $J=2-1$ & $122 \times 115$ & 1.9 & 0.35 & \makecell{DSHARP} \\
LkCa 15 & GO-1282 & 224 & $50 \times 50$ & 0.004 & \makecell{\citet{Long2022_lkca15}} & $J=3-2$ & $150 \times 150$ & 2.79 & 0.1 & \makecell{exoALMA} \\
MY Lup & GO-1584 & 239 & $43 \times 43$ & 0.016 & \makecell{DSHARP} & $J=2-1$ & $100 \times 82$ & 1.08 & 0.35 & \makecell{DSHARP} \\
PDS 70 & GO-1282 & 232 & $115 \times 89$ & 0.017 & \makecell{2019.1.01619.S} & $J=2-1$ & $148 \times 118$ & 1.18 & 0.4 & \makecell{2019.1.01619.S} \\
RNO 90 & GO-1282 & 402 & $134 \times 106$ & 0.239 & \makecell{2015.1.00847.S} & $J=3-2$ & $343 \times 238$ & 8.77 & 0.4 & \makecell{2015.1.00168.S} \\
RU Lup & GO-1584 & 239 & $24 \times 24$ & 0.021 & \makecell{DSHARP} & $J=2-1$ & $95 \times 83$ & 1.12 & 0.35 & \makecell{DSHARP} \\
RW Aur & GO-1282 & 224 & $30 \times 17$ & 0.009 & \makecell{\citet{Kurtovic2024}} & $J=2-1$ & $100 \times 100$ & 0.38 & 1.3 & \makecell{\citet{Kurtovic2024}} \\
RY Lup & GO-1640 & 222 & $138 \times 120$ & 0.034 & \makecell{2017.1.00449.S} & $J=2-1$ & $194 \times 166$ & 3.23 & 0.4 & \makecell{2017.1.00449.S} \\
SR 21 & GO-1282 & 217 & $39 \times 30$ & 0.012 & \makecell{\citet{Yang2023}} & $J=2-1$ & $111 \times 96$ & 1.42 & 0.4 & \makecell{2018.1.00689.S} \\
SR 4 & GO-1584 & 239 & $34 \times 34$ & 0.025 & \makecell{DSHARP} & $J=2-1$ & $111 \times 87$ & 1.06 & 0.35 & \makecell{DSHARP} \\
SY Cha & GO-1282 & 338 & $129 \times 96$ & 0.031 & \makecell{exoALMA} & $J=3-2$ & $150 \times 150$ & 3.4 & 0.1 & \makecell{exoALMA} \\
Sz 129 & GO-1584 & 239 & $43 \times 31$ & 0.015 & \makecell{DSHARP} & $J=2-1$ & $110 \times 83$ & 0.9 & 0.35 & \makecell{DSHARP} \\
Sz 98 & GO-1282 & 225 & $94 \times 70$ & 0.061 & \makecell{2015.1.00222.S\\2018.1.01458.S} & $J=2-1$ & $196 \times 175$ & 5.59 & 0.4 & \makecell{2015.1.00222.S\\2018.1.01458.S} \\
TW Cha & GO-1549 & 225 & $121 \times 85$ & 0.05 & \makecell{2021.1.00854.S} & $J=2-1$ & $289 \times 155$ & 4.16 & 0.5 & \makecell{2021.1.00854.S} \\
TW Hya & GO-1282 & 290 & $37 \times 25$ & 0.02 & \makecell{\citet{Andrews2016}} & $J=3-2$ & $139 \times 131$ & 3.44 & 0.25 & \makecell{\citet{Huang2018_twhya}} \\
V1094 Sco & GO-1282 & 239 & $62 \times 53$ & 0.027 & \makecell{2017.1.01167.S} & $J=2-1$ & $161 \times 161$ & 8.36 & 0.1 & \makecell{2017.1.01167.S} \\
VW Cha & GO-1282 & 226 & $76 \times 52$ & 0.059 & \makecell{2021.1.00854.S} & $J=2-1$ & $221 \times 137$ & 3.5 & 0.5 & \makecell{2021.1.00854.S} \\
VZ Cha & GO-1549 & 225 & $80 \times 57$ & 0.039 & \makecell{2021.1.00854.S} & $J=2-1$ & $217 \times 164$ & 1.53 & 1.0 & \makecell{2021.1.00854.S} \\
WaOph 6 & GO-1282 & 239 & $58 \times 53$ & 0.017 & \makecell{DSHARP} & $J=2-1$ & $126 \times 115$ & 1.32 & 0.35 & \makecell{DSHARP} \\
WSB 52 & GO-1584 & 239 & $33 \times 26$ & 0.016 & \makecell{DSHARP} & $J=2-1$ & $187 \times 159$ & 1.25 & 0.4 & \makecell{DSHARP} \\
WX Cha & GO-1282 & 226 & $79 \times 58$ & 0.074 & \makecell{2021.1.00854.S} & $J=2-1$ & $200 \times 123$ & 4.55 & 0.5 & \makecell{2021.1.00854.S} \\
XX Cha & GO-1282 & 226 & $67 \times 40$ & 0.028 & \makecell{2021.1.00854.S} & $J=2-1$ & $152 \times 99$ & 2.23 & 0.5 & \makecell{2021.1.00854.S} \\
\enddata
\tablecomments{The ``REF" columns list the ALMA program IDs corresponding to the data used for each target or the references (DSHARP=\citealt{Andrews2018}, exoALMA=\citealt{Teague2025}) where data products were adopted directly. Column (11) gives the velocity channel width of the CO data cube.}
\end{deluxetable*}

\section{Visibility plane fitting for disks with new ALMA data} \label{sec:appendix-uvfit}
To characterize the dust emission and disk geometry, we performed model fitting directly in the visibility domain. We adopted the ``Nuker'' surface brightness profile due to its flexibility in describing diverse disk morphologies (e.g., \citealt{Tripathi2017}), expressed as
\begin{equation} \label{eq1}
    I_{\nu}(r) \propto \left(\frac{r}{r_t} \right)^{-\gamma} \left[ 1 + \left (\frac{r}{r_t} \right) ^{\alpha} \right ] ^{(\gamma - \beta)/\alpha},
\end{equation}
where $r_t$ is the transition radius, $\alpha$ controls the sharpness of the transition, and $\gamma$ and $\beta$ describe the inner and outer slopes, respectively. The profile is then normalized by the total flux density $F_{\nu}$, corresponding to the integral over all radii.
Model visibilities were computed via Hankel transformation \citep{Pearson1999} and sampled at the same spatial frequencies as the observations, including corrections for positional offsets from the phase center (dRA, dDec), disk inclination ($i$), and rotation by a sky-plane position angle (PA). This results in a total of 9 free parameters.
The comparison of model and data visibilities used a Gaussian likelihood $\mathcal{L}\propto \exp(-\chi^2/2)$, where $\chi^{2}= \sum \left| V_{\mathrm{obs}}(u_{k}, v_{k})-V_{\mathrm{mod}}(u_{k}, v_{k}) \right|^{2} w_{k}$, with $w_{k}$ denoting the observed visibility weights. Uniform priors were assumed for all parameters, and the parameter space was explored using a Markov Chain Monte Carlo (MCMC) approach (\texttt{emcee}, \citealt{ForemanMackey2013}) with 45 walkers and 5000 steps.
Final adopted parameters (Table~\ref{tab:uvfit}) are the medians of the posterior distributions from the last 1000 steps, with uncertainties from the 16th and 84th percentiles. For the faint disk around HD~39539, a point source model ($F_{\nu}$, dRA, dDec) is sufficient to describe the observed emission. Disk properties of the close binary DF Tau can be found in \citet{Grant2024_DFTau}.

\begin{figure}[th!]
    \centering
    \includegraphics[width=\linewidth]{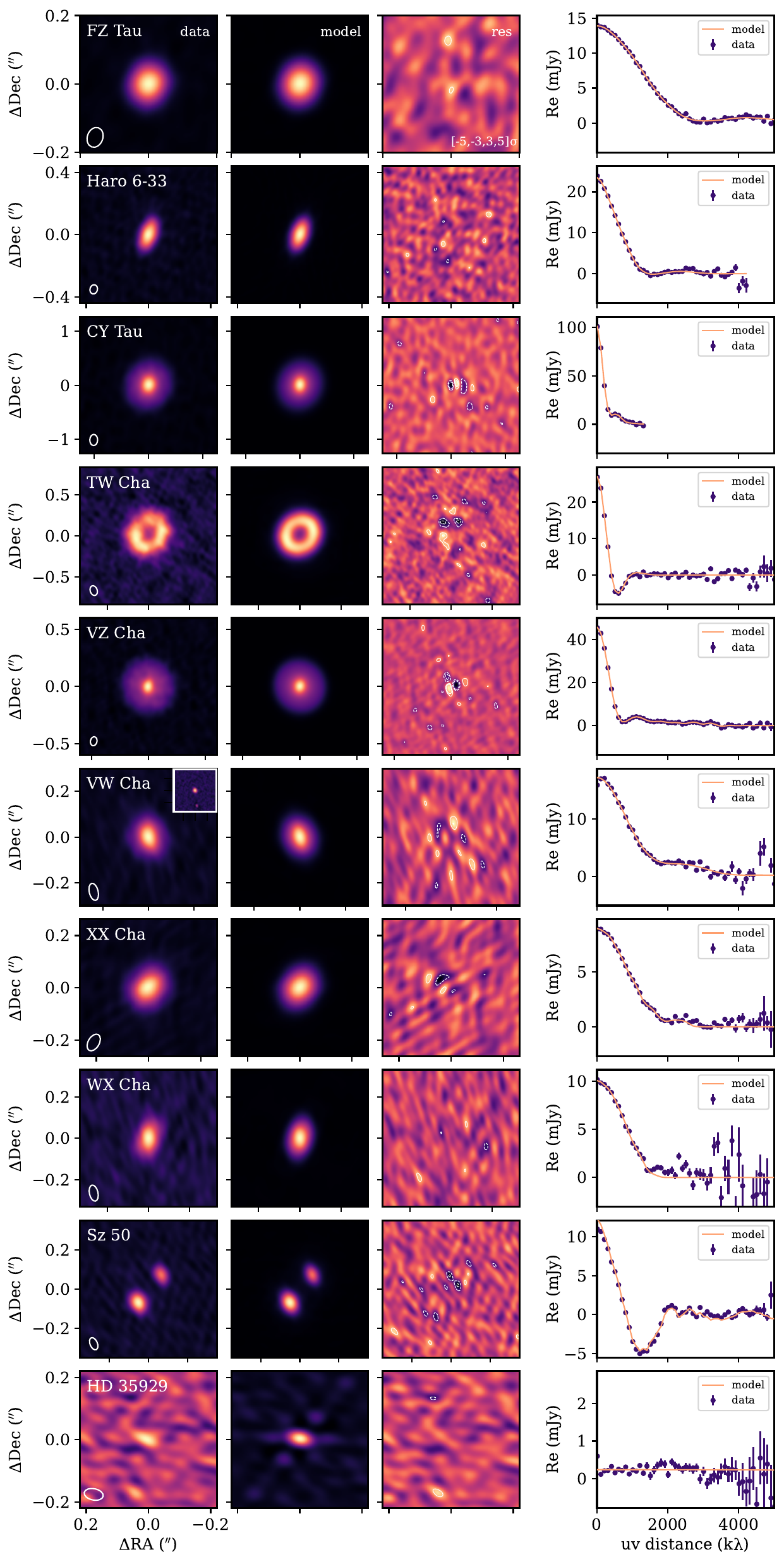}
    \caption{Comparison between our parametric intensity models and the observations. The first three columns show the observed, modeled, and residual images, generated using identical \texttt{TCLEAN} parameters. The insert for VW~Cha highlights the faint companion disk located $\sim0\farcs65$ to the south. The last column presents the deprojected and binned observed visibilities compared with the model visibilities, except for Sz~50 and HD~35929, whose visibilities were not deprojected.}
    \label{fig:cont_newdata}
\end{figure}

Figure~\ref{fig:cont_newdata} compares the best-fit models with the observations. The axisymmetric models provide overall good fits, with residuals typically within 3--5$\sigma$. Only TW~Cha exhibits clear substructures, while other disks, such as CY~Tau and VZ~Cha, may host small-scale features that remain unresolved at the current angular resolution. Our data reveal Sz~50 as a close binary with a projected separation of $\sim0\farcs2$, and identify weak emission from the southern companion of VW Cha (at $\sim0\farcs65$).

\begin{deluxetable*}{lrrrrrrrrr}
\tabletypesize{\scriptsize}
\tablecaption{Nuker Profile Fitting Result\label{tab:uvfit}}
\tablewidth{0pt}
\tablehead{
\colhead{Name} & \colhead{$F_{\nu}$} & \colhead{$r_t$} & \colhead{$log\alpha$} & \colhead{$\beta$} & \colhead{$\gamma$} & \colhead{$i$} & \colhead{$PA$} & \colhead{$\Delta$RA} & \colhead{$\Delta$Dec} \\
\colhead{} & \colhead{(mJy)} & \colhead{(arcsec)} & \colhead{} & \colhead{} & \colhead{} & \colhead{($\deg$)} & \colhead{($\deg$)} & \colhead{(mas)} & \colhead{(mas)} \\
}
\colnumbers
\startdata
FZ Tau & $13.86 \pm 0.02$ & $0.054 \pm 0.001$ & $1.26^{+0.23}_{-0.10}$ & $9.30^{+0.45}_{-0.62}$ & $0.64 \pm 0.01$ & $13.4 \pm 1.0$ & $169.3 \pm 4.2$ & $-0.19 \pm 0.06$ & $0.19 \pm 0.06$ \\
Haro 6-33 & $23.37 \pm 0.04$ & $0.088 \pm 0.001$ & $1.82^{+0.12}_{-0.17}$ & $4.67 \pm 0.06$ & $0.44 \pm 0.01$ & $60.1 \pm 0.2$ & $161.0 \pm 0.2$ & $-0.26 \pm 0.09$ & $0.41 \pm 0.12$ \\
CY Tau & $105.83 \pm 0.10$ & $0.453 \pm 0.007$ & $1.07 \pm 0.04$ & $8.39 \pm 0.36$ & $0.86 \pm 0.00$ & $26.1 \pm 0.2$ & $153.1 \pm 0.5$ & $-0.11 \pm 0.17$ & $0.05 \pm 0.20$ \\
TW Cha & $27.55 \pm 0.05$ & $0.245 \pm 0.002$ & $0.82 \pm 0.03$ & $9.88^{+0.09}_{-0.20}$ & $-1.65 \pm 0.11$ & $29.6 \pm 0.5$ & $129.9 \pm 1.0$ & $1.87 \pm 0.43$ & $16.00 \pm 0.48$ \\
VZ Cha & $46.46 \pm 0.04$ & $0.219 \pm 0.000$ & $1.92^{+0.06}_{-0.11}$ & $9.97^{+0.02}_{-0.04}$ & $0.74 \pm 0.00$ & $18.8 \pm 0.4$ & $31.3 \pm 1.4$ & $0.47 \pm 0.13$ & $-0.13 \pm 0.18$ \\
VW Cha & $17.32 \pm 0.03$ & $0.089 \pm 0.002$ & $1.06^{+0.22}_{-0.11}$ & $9.80^{+0.15}_{-0.31}$ & $0.97 \pm 0.02$ & $26.6 \pm 1.3$ & $36.0 \pm 3.2$ & $0.37 \pm 0.16$ & $0.22 \pm 0.21$ \\
XX Cha & $8.95 \pm 0.02$ & $0.084 \pm 0.002$ & $0.75 \pm 0.05$ & $9.56^{+0.31}_{-0.59}$ & $0.60 \pm 0.04$ & $27.0 \pm 1.1$ & $138.6 \pm 2.6$ & $0.29 \pm 0.17$ & $-0.05 \pm 0.21$ \\
WX Cha & $10.08 \pm 0.03$ & $0.088 \pm 0.003$ & $1.50 \pm 0.36$ & $9.32^{+0.47}_{-0.87}$ & $0.78 \pm 0.04$ & $53.7 \pm 0.9$ & $155.5 \pm 1.6$ & $0.08 \pm 0.30$ & $0.35 \pm 0.43$ \\
Sz 50 A & $7.76 \pm 0.05$ & $0.019 \pm 0.001$ & $1.67 \pm 0.26$ & $3.11 \pm 0.04$ & $-2.59 \pm 0.50$ & $22.2 \pm 2.9$ & $50.2 \pm 8.7$ & $51.29 \pm 0.18$ & $-69.88 \pm 0.21$ \\
Sz 50 B & $5.10 \pm 0.11$ & $0.011 \pm 0.001$ & $1.52 \pm 0.35$ & $2.52 \pm 0.04$ & $-2.18 \pm 0.77$ & $15.1 \pm 7.0$ & $27.5^{+46.2}_{-99.7}$ & $-62.90 \pm 0.29$ & $70.33 \pm 0.34$ \\
HD 35929 & $0.24 \pm 0.01$ & & & & & & & $-0.32\pm2.42$& $1.95\pm1.67$ \\
\enddata
\tablecomments{The posterior distribution of $\beta$ for TW Cha, VZ Cha, VW Cha, XX Cha, and WX Cha has a preference toward the upper bound of the prior on $\beta$.}
\end{deluxetable*}

\section{Disk size measure} \label{sec:appendix-disk-size}
\begin{figure*}[t!]
    \centering
    \includegraphics[width=\linewidth]{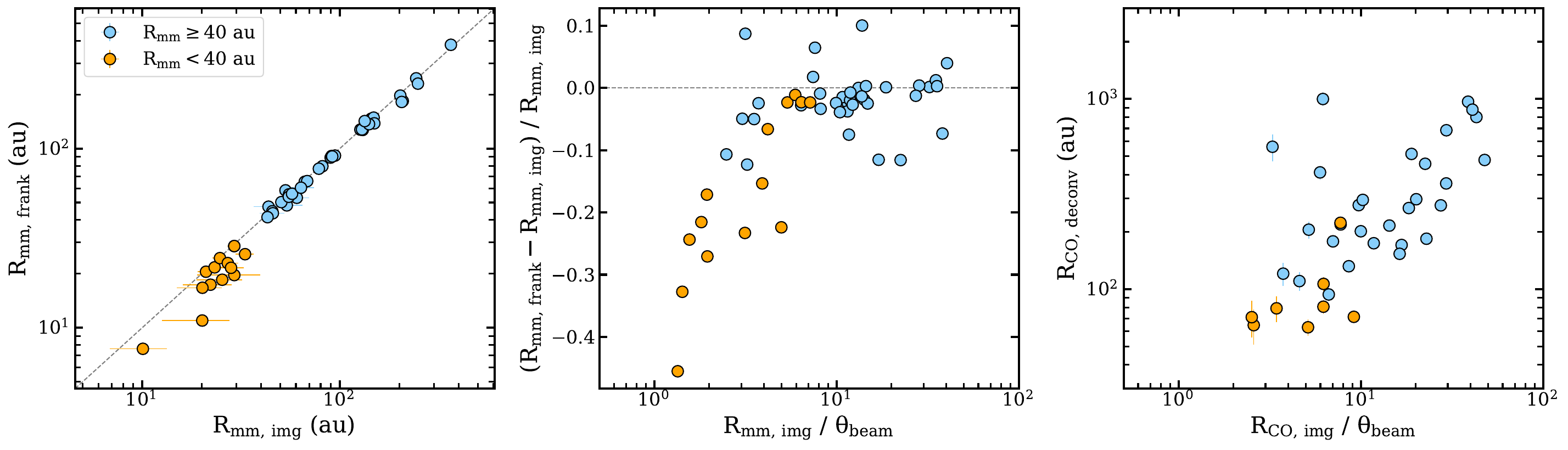}
    \caption{\emph{Left:} Millimeter continuum disk sizes measured from the image plane vs.\ those derived from \texttt{frank} visibility-plane fitting, with the grey dashed line indicating the 1:1 correspondence. \emph{Middle:} Fractional difference between the two continuum size measurements as a function of the number of resolution elements across the disk. \emph{Right:} Beam-deconvolved CO gas disk size as a function of the number of CO resolution elements.}
    \label{fig:appendix-size}
\end{figure*}

Figure~\ref{fig:appendix-size} compares the image-plane and \texttt{frank} continuum sizes across our sample. The two agree well for large disks, while \texttt{frank} returns systematically smaller sizes for the most compact, marginally resolved sources, as expected when beam convolution broadens the radial intensity profile and biases the image-plane $R_{90\%}$ high. The CO sizes are far less affected, as all disks with reliable CO measurements are well resolved and the beam deconvolution introduces only a minor correction.

\bibliography{sample631}{}
\bibliographystyle{aasjournal}
\end{CJK*}
\end{document}